\documentstyle[12pt]{article}

\catcode`\@=11

\@addtoreset{equation}{section}
\def\theequation{\arabic{section}.\arabic{equation}}

\def\@normalsize{\@setsize\normalsize{15pt}\xiipt\@xiipt
\abovedisplayskip 14pt plus3pt minus3pt%
\belowdisplayskip \abovedisplayskip
\abovedisplayshortskip  \z@ plus3pt%
\belowdisplayshortskip  7pt plus3.5pt minus0pt}

\def\small{\@setsize\small{13.6pt}\xipt\@xipt
\abovedisplayskip 13pt plus3pt minus3pt%
\belowdisplayskip \abovedisplayskip
\abovedisplayshortskip  \z@ plus3pt%
\belowdisplayshortskip  7pt plus3.5pt minus0pt
\def\@listi{\parsep 4.5pt plus 2pt minus 1pt
            \itemsep \parsep
            \topsep 9pt plus 3pt minus 3pt}}

\@twosidetrue

\relax

\catcode`@=12

\catcode`\@=11

\def\section{\@startsection{section}{1}{\z@}{3.5ex plus 1ex minus
   .2ex}{2.3ex plus .2ex}{\large\bf}}

\def\thesection{\Roman{section}.}

\def\appendix{\setcounter{section}{0}
        \def\thesection{Appendix }
        \def\theequation{\Alph{section}.\arabic{equation}}}

\def\ps@headings{\def\@oddfoot{}\def\@evenfoot{}
\def\@oddhead{\hbox{}\hfill
        \makebox[.5\textwidth]{\raggedright\ignorespaces --\thepage{}--
        \hfill {}}}
\def\@oddhead{\hbox{}\hfill --\thepage{}-- \hfill
        {}}
\def\@evenhead{\@oddhead}
\def\subsectionmark##1{\markboth{##1}{}}
}

\ps@headings

\catcode`\@=12

\relax

\def\figcap{\section*{Figure Captions\markboth
        {FIGURECAPTIONS}{FIGURECAPTIONS}}\list
        {Fig. \arabic{enumi}:\hfill}{\settowidth\labelwidth{Fig. 999:}
        \leftmargin\labelwidth
        \advance\leftmargin\labelsep\usecounter{enumi}}}
 \relax
\def\tablecap{\section*{Table Captions\markboth
        {TABLECAPTIONS}{TABLECAPTIONS}}\list
        {Table \arabic{enumi}:\hfill}{\settowidth\labelwidth{Table 999:}
        \leftmargin\labelwidth
        \advance\leftmargin\labelsep\usecounter{enumi}}}
 \relax
\def\reflist{\section*{References\markboth
        {REFLIST}{REFLIST}}\list
        {[\arabic{enumi}]\hfill}{\settowidth\labelwidth{[999]}
        \leftmargin\labelwidth
        \advance\leftmargin\labelsep\usecounter{enumi}}}
 \relax

\catcode`\@=11

\def\ps@headings{\def\@oddfoot{}\def\@evenfoot{}
\def\@oddhead{\hbox{}\hfill
        \makebox[.5\textwidth]{\raggedright\ignorespaces --\thepage{}--
        \hfill {}}}
\def\@evenhead{\@oddhead}
\def\subsectionmark##1{\markboth{##1}{}}
}

\ps@headings

\relax

\newskip\humongous \humongous=0pt plus 1000pt minus 1000pt

\newif\ifdtup

\def\beq{\begin{equation}}
\def\eeq{\end{equation}}

\def\beqn{\begin{eqnarray}}
\def\eeqn{\end{eqnarray}}
\relax

\def\G2{{\; \rm GeV/}c^2}
\def\G{\; \rm GeV}

\def\dotx{\dotx{\dot\overline{x}}}

\relax

\newcommand {\al}{\alpha}
\newcommand {\be}{\beta}
\newcommand {\de}{\delta}

\newcommand {\ep}{\epsilon}

\newcommand {\ran}{\rangle}
\newcommand {\lan}{\langle}

\newcommand {\lam}{\lambda}
\newcommand {\te}{\theta}
\newcommand {\sig}{\sigma}
\newcommand {\Sig}{\Sigma}
\newcommand {\bte}{\bar{\theta}}
\newcommand {\bla}{\bar{\lambda}}
\newcommand {\bsi}{\bar{\sigma}}
\newcommand {\bet}{\bar{\eta}}
\newcommand {\dal}{\dot{\alpha}}
\newcommand {\dbe}{\dot{\beta}}
\newcommand {\pal}{\partial}
\newcommand {\non}{\nonumber}
\newcommand {\1}[1]{\frac{1}{#1}}
\newcommand {\im}[1]{\frac{i}{#1}}
\newcommand {\2}[1]{\frac{\delta}{\delta #1}}
\newcommand{\yp}{y^{\prime}}
\newcommand{\zp}{z^{\prime}}
\newcommand{\bta}{\bar{\tau}}
\newcommand{\bw}{\bar{w}}
\newcommand{\dw}{\dot{w}}
\newcommand{\cA}{{\cal A}}
\newcommand{\cP}{{\cal P}}
\newcommand{\cL}{{\cal L}}
\newcommand{\rD}{{\rm D}}
\newcommand{\rV}{{\rm V}}
\newcommand{\brD}{\bar{\rm D}}
\newcommand{\rDp}{{\rm D}^{\prime}}
\newcommand{\brDp}{\bar{\rm D}^{\prime}}

\begin{document}

\begin{titlepage}
\begin{flushright}
       {\normalsize    OU-HET 259 \\  hep-th 9703***, \\
                March,~1997  }
\end{flushright}
\vfill
\begin{center}
  {\large \bf   Schwinger-Dyson Equation for
      \\  Supersymmetric Yang-Mills Theory: \\
      Manifestly Supersymmetric Form }\footnote{This work is
 supported in part by  Grant-in-Aid for  Scientific Research
(07640403)
from
 the Ministry of Education, Science and Culture, Japan.}

\vfill

         {\bf H.~Itoyama}  \\
            and \\
              {\bf H.~Takashino}
\vfill
        Department of Physics,\\
        Graduate School of Science, Osaka University,\\
        Toyonaka, Osaka, 560 Japan\\
\end{center}
\vfill

\begin{abstract}
   We study our Schwinger-Dyson equation as well as the large $N_{c}$ loop
equation for supersymmetric Yang-Mills theory in four dimensions by the $N=1$
superspace Wilson-loop variable.  We are successful in deriving a new
manifestly supersymmetric form  in which a loop splitting and joining are
represented by a manifestly supersymmetric as well as supergauge invariant
operation in superspace.  This is found to be a natural extension from the
abelian case. We solve the equation to leading order in
perturbation theory or equivalently in the linearized approximation,
obtaining a desirable nontrivial answer. The super Wilson-loop variable can
be represented as the system of one-dimensional fermion along the loop
coupled minimally to the original theory. One-loop renormalization of the
one-point Wilson-loop average  is explicitly carried out, exploiting this
property. The picture of string dynamics obtained is briefly discussed.

\end{abstract}
\vfill
\end{titlepage}

\section {Introduction}

Intensive efforts have recently been directed to both supersymmetric gauge
 theories and superstring unification.  Main motivation is that we must be
 able to discuss nonperturbative phenomena for these intelligibly  before
  we confront ourselves to reality.
  In one side we have rich phenomena which supersymmetric gauge theories
 offer. These will become
  hints to how string theory should be formulated in the other.
 Connection between superstrings and gauge theories is thus beginning to play
 important roles in recent activities and will continue to progress in
 various directions.
  It has been known, in a related context, that the
 formulation of gauge theories as Schwinger-Dyson equations which exploit
 Wilson-loops as  basic variables provides a natural framework  to discuss
  nonperturbative phenomena such as confinement and dynamical symmetry
 breaking. At the same time, it exhibits string interactions contained in
 gauge theories as loop dynamics~\cite{MM,Mig,Pol}.  We wish to provide a 
 significant step to this approach in this paper, which is relevant to 
 current issues.

 In the previous paper~\cite{IT}, we initiated the (super)-gauge invariant 
 formulation of supersymmetric gauge theories as Schwinger-Dyson equations.
 We proposed to use the Wilson-loop having the connection one-form 
  in $N=1$ superspace  as fundamental variable. 
   We will refer to this variable as super 
 Wilson-loop~\cite{Gates}-\cite{GGRS} in this paper.
  This setup owes  some of the good properties  to  the geometric formulation
 \cite{WB,So} 
 of supersymmetric gauge theories on $N=1$ superspace, which is by now
 well understood in the literature \footnote{ There were several attempts
 in the earlier literature~\cite{GJ,MakMed},
 trying to derive supersymmetric loop equations.
  These were hampered by the lack of this geometrical formulation. }.
 
 We have been successful in deriving the Schwinger-Dyson equation as well as
  the large $N_{c}$ loop equation in a closed form consisting
 of the super Wilson-loop alone.
 The (super)-area derivative acting on the super Wilson-loop, which
 includes the Grassmannian directions, is found to be a central 
  geometric operation to our equation.
  The final form of our equation looked, however, complicated.
 Although it is closed with respect to the super Wilson-loop, it was not clear
 how we could make further progress  with this.  The basic reason for this
  complexity comes from the fact that we have to work with the Wess-Zumino
gauge in order to carry out algebras in intermediate steps.
 Manifest supersymmetry is lost this way.  One of the upshots of the present
 paper is that we are able to overcome
  this difficulty and  to  put the equation in the manifestly
 supergauge invariant form.
 
 Another aspect of the super Wilson-loop is  concerned with the question
 of the removal of infinities.  The super Wilson-loop is a composite operator
 and contains infinite number of insertions of gauginos and  the auxiliary
 ${\rm D}$ field into the ordinary Wilson-loop along the curve.
 One may think that renormalization of this composite operator is 
 prohibitively complicated.  We will show that this is not the case.
 The renormalization of the super Wilson-loop is no more complicated
  than that of the ordinary Wilson-loop~\cite{IH}-\cite{Ao}.

 In the next section, we introduce   the superspace one-form from
the vector superfield ${\rm V}$ which is followed by the construction of the
 super Wilson-loop. We list several important properties of the super-Wilson
loop.  In section three,  we first derive the Schwinger-Dyson equation  in the
 abelian case. Although this is a free theory, it provides us with
 an important insight into how we ``gauge-unfix''  the equation in subsequent
 sections. In section four,  we present  the derivation of our equation in
 the Wess-Zumino gauge.  The equation we obtained  is (\ref{eq.SS-D})
\beqn
& &\1{8g^2}\ep_{\al\be}\bar{\sig}^{a\dot{\al}\be}{\rm D}^{\al}
   \2{\Sig^{a\dot{\al}}(\zp)}\lan tr W_S[C] \ran  \non\\
&=&-2\oint \{-\im{2}dy^a(\eta-\te)\sig_a(\bet-\bte)
-\1{2}dy^a\de(\eta-\te)\bte\bsi^b\sig_a\bet\pal_b\non\\
& &\hspace*{0.7cm}+\bet^2\left(d\eta(\eta-\te)
+id\eta\sig^a\bte\de(\eta-\te)\pal_a\right) \}
\lan tr\,W_S[C_{z^{\prime}z}] \de^4(y-y^{\prime})
                tr\,W_S[C_{zz^{\prime}}]\ran \non\\
& &+2\oint\1{8}
\bet^2\de(\eta-\te)\de^4(y-y^{\prime})dy^a(\bsi^b\sig_a\bte)^{\dal}\non\\
& &\times \left\{\lan(\2{\Sig^{b\dal}(z)} tr\,W_S[C_{z^{\prime}z}])
                       tr\,W_S[C_{zz^{\prime}}]\ran
                    -\lan tr\,W_S[C_{z^{\prime}z}]
            (\2{\Sig^{b\dal}(z)}tr\, W_S[C_{zz^{\prime}}])\ran \right\}\,.
       (\ref{eq.SS-D})\non
\eeqn
 Only the ordinary gauge invariance is kept intact. 

  Section five deals with our advancement to the manifestly supersymmetric
 form.  Being motivated by the abelian case, we are able to find how to make
 (\ref{eq.SS-D}) into manifestly supergauge invariant. This amounts
  to recovering the Wess-Zumino gauge volume, which is thrown away upon
 partial gauge fixing.  The final form of our equation reads
\beqn
\hspace*{3.5cm}& &\1{8g^2}\ep_{\al\be}\bsi^{a\dal\be}{\rm D}^{\al}
\2{\Sig^{a\dal}(\zp)}
\lan tr\,W_S[C]\ran  \non\\
&=&\oint{\cal D}^{\al}_f {\rm D}_{\al}\de(z-\zp)
\lan tr\,W_S[C_{\zp z}]tr\,W_S[C_{z\zp}] \ran \,,\non\\
& &{\cal D}_f^{\al}
=-e^a\frac{i}{4}\bar\sig_a^{\dal\al}\bar{\rm D}_{\eta\dot{\al}}+e^{\al}\,,
\non\\
& &\lan\,\cdots\,\ran =\int[d{\rm V}]\,{\rm e}^{i{\rm S}_{SYM}}\,\cdots\,.
      \hspace*{5cm} (\ref{eq.SS-Dfinal})\non
\eeqn
We briefly discuss the picture of string dynamics obtained from the sequence 
of the Schwinger-Dyson equations beginning with this one.
 In section six, we solve the equation in the abelian 
case.  This is equivalent to solving  the full-fledged nonabelian
 case in the linearized approximation  or to the leading order in
 perturbation theory.  In section seven, we deal with the problem
 of removing infinities  and renormalization.  Representing the 
 super Wilson-loop as the first quantized system of one-dimensional 
 fermion along the loop,
  this problem becomes   that for ${\cal L}_{SYM}+ {\cal L}_{path}$. This
 can then be handled in the standard fashion.  We carry out the one-loop
 renormalization explicitly and  the infinities are shown to cancel
 by local counterterms.  In the final section, we discuss on a few points.
  In the first two of  three appendix, we give some pedagogical details on
  the area derivative and the Migdal-Makeenko equation.
  In the final appendix, we summarize the basic formulas
  on the connection one-form in superspace and the vector superfield
 known in the literature and used frequently in the text.
  Unless written explicitly, we follow the notation of \cite{WB}.

%
\section{Supersymmetric Wilson-loop}
Let us consider
 the supersymmetric Schwinger-Dyson equation using the Wilson-loop variable.
 In parallel to the ordinary path ordered exponential of
  the one-form
 $dx^av_a$ on spacetime $x$, we define the path ordered
 exponential of the one-form ${\cal A}$ on superspace
 $z$ \cite{Gates}-\cite{GGRS}, \cite{GJ}.
 From (3.77) and (\ref{eq.sol}), ${\cal A}$ is written in terms of the vector 
superfield ${\rm V}$ as
\beqn
{\cal A}=e^{\rm A}{\cal A}_{\rm A}    
 &=& (dx^a-id\eta\sig^a\bar{\eta}+i\eta\sig^ad\bar{\eta}){\cal A}_a  
+d\eta^{\al}{\cal A}_{\al} \,,   \non\\
{\cal A}_{\alpha}
&=&-{\rm e}^{-{\rm V}}{\rm D}_{\alpha}{\rm e}^{\rm V} \,,\nonumber\\
\bar{{\cal A}}_{\dot{\alpha}}  &=&0 \,,\nonumber\\
{\cal A}_a
&=&\frac{i}{4}\bar{\sigma}^{\dot{\alpha}\alpha}_a
   \bar{{\rm D}}_{\dot{\alpha}}{\rm e}^{-{\rm V}}{\rm D}_{\alpha}
                                            {\rm e}^{\rm V} \,,   \label{eq.Ax}
\eeqn
where $(x,\eta,\bar{\eta})$ denotes the superspace coordinates.
In terms of $(y=x+i\eta\sig^a\bar{\eta},\eta,\bar{\eta})$,   (\ref{eq.Ax})
 becomes 
\beqn
{\cal A}=(dy^a-2id\eta\sig^a\bar{\eta}){\cal A}_a+d\eta^{\al}{\cal A}_{\al}\,.
                                                                  \label{eq.Ay}
\eeqn

We introduce the curve $C_{z_1z_2}$ on superspace, using one 
dimensional parameter $t$:
\beqn
C_{z_1z_2}:z^M
&=&z^M(t)  \non\\
&=&(x(t),\eta(t),\bar{\eta}(t)),  \non\\
0\le t\le 1,& &z(0)=z_2,\,\,\,\,z(1)=z_1.
\eeqn
One can think of the Grassmannian coordinates $\eta(t)$ or $\bar{\eta}(t)$
 as being defined by the infinite number of 
Grassmannian parameters via Taylor expansion
\beqn 
\eta(t)
&=&\sum_{n=0}^{\infty}\frac{1}{n!}\eta^{(n)}t^n \,,\non\\
\bar{\eta}(t)
&=&\sum_{n=0}^{\infty}\frac{1}{n!}\bar{\eta}^{(n)}t^n \,.\non
\eeqn
Supersymmetric path-ordered exponential 
$W_S[C_{z_1z_2}]$ is given by
\beqn
W_S[C_{z_1z_2}]
&=&\sum_{n=0}^{\infty}\int_0^1{\cal A}(z(t_1))\int_0^{t_1}{\cal A}(z(t_2))
\cdots\int_0^{t_{n-1}}{\cal A}(z(t_n))              \non\\
&\equiv&P\exp\int_{z_2}^{z_1}{\cal A}(z)\,\,,                \label{eq.defsW}\\
{\cal A}(z(t))
&=&dt\frac{dz^M}{dt}{\cal A}_M(z(t))     \non\\
&=&dt\{\frac{dx^a}{dt}-i\frac{d\eta}{dt}\sig^a\bar{\eta}
   +i\eta\sig^a\frac{d\bar{\eta}}{dt}\}{\cal A}_a  \non\\
& &+dt\frac{d\eta^{\al}}{dt}{\cal A}_{\al}\,.                  \label{SWA}
\eeqn
In terms of $(y,\eta,\bar{\eta})$
\beqn
{\cal A}(z(t))=dt\{\frac{dy^a}{dt}
               -2i\frac{d\eta}{dt}\sig^a\bar{\eta}\}{\cal A}_a 
               +dt\frac{d\eta^{\al}}{dt}{\cal A}_{\al}\,.
\eeqn
Here the capital $P$ denotes the path-ordered product in superspace.
 Since the lowest component of 
${\cal A}_a$ is $-\frac{i}{2}v_a$,  $W_S$ becomes 
  the ordinary path-ordered exponential $W$ if we take $\eta=\bar{\eta}=0$.

The $W_S[C_{z_1z_2}]$ can be characterized by
\beqn
\frac{d\phi(z(t))}{dt}&=&{\cal A}(z(t))\phi(z(t))\,,  \non\\
\phi(z_1)&=&W_S[C_{z_1z_2}]\phi(z_2)\,,                        \label{eq.sdiff}
\eeqn
where $\phi(z)$ denotes a certain superfield belonging to a representation 
of the gauge group.
 Note that the one-form ${\cal A}$ is given by (\ref{eq.Ax}) and is expressed
 with the vector superfield ${\rm V}$ defined by (\ref{eq.vector}).
 It contains not only
the usual Yang-Mills fields $v_a$ but also its partners $\lam,\bla$ and
${\rm D}$. 
 From the definition of $W_S[C_{z_1z_2}]$, namely (\ref{eq.defsW}), 
 we see that 
$W_S[C_{z_1z_2}]$ has infinite number of insertions 
$\lam,\bla$ and ${\rm D}$ along the curve $C_{z_1z_2}$.  The fields 
$\lam,\bla$ and ${\rm D}$ all  belong to the adjoint representation
  and should be treated in the same way as $v_a$.  The $W_S[C_{z_1z_2}]$ is
natural from that perspective, which becomes most evident in the large $N_{c}$
 limit.
The operator $W_S$ possesses the properties similar to those of 
  the ordinary path-ordered exponential $W$.

(A) $W_S$ has reparametrization invariance for curves in superspace. 

(B) For the addition of two curves,
$W_S[C_{z_1z_2}]=W_S[C_{z_1z}]W_S[C_{zz_2}]=W_S[C_{z_1 z}+C_{z z_1}]$.

(C) For the deformation of the curve $C_{z_1z_2}$:$z(t) \to z(t)+\de z(t)$,
 we obtain
\beqn 
\de W_S[C_{z_1z_2}]
&=&\int_{z_2}^{z_1}dz^M\de z^NW_S[C_{z_1z}]{\cal F}_{NM}(z)W_S[C_{zz_2}] \non\\
& &+\{\de z_1^M{\cal A}_M(z_1)W_S[C_{z_1z_2}]
   -W_S[C_{z_1z_2}]\de z_2^M{\cal A}_M \}                \label{eq.sconvo}
\eeqn
where $\de z_1=\de z(1),\,\,\de z_2=\de z(0)$ and ${\cal F}_{NM}(z)$
 having Einstein indices are the components of the field strength ${\cal F}$
\beqn
{\cal F}
&=& d{\cal A}+{\cal A}{\cal A} = \frac{1}{2} dz^M dz^N {\cal F}_{NM}\non\\
&=&\frac{1}{2} dz^M dz^N \{\pal_N {\cal A}_M-(-)^{|N||M|}\pal_M{\cal A}_N
                                                                 \non\\
& &\hspace*{2cm}-{\cal A}_N{\cal A}_M+(-)^{|N||M|}{\cal A}_M{\cal A}_N \}\,.
\eeqn
From (\ref{eq.sconvo}), we can define the end point 
derivatives of $W_S[C_{z_1z_2}]$:
\beqn
\frac{\pal}{\pal z_1^M}W_S[C_{z_1z_2}]
={\cal A}_M(z_1)W_S[C_{z_1z_2}]\;\;, \;\;
\frac{\pal}{\pal z_2^M}W_S[C_{z_1z_2}] = 
-W_S[C_{z_1z_2}]{\cal A}_M(z_2)  \,.    \label{eq.sderivative2}
\eeqn
  Hitting with the derivative ${\rm D}_A$ having flat indices ( see
 (\ref{eq.113})), we obtain 
\beqn
{\rm D}_A^{(z_{1})}W_S[C_{z_1z_2}]
={\cal A}_A(z_1)W_S[C_{z_1z_2}]    \,,\non\\
{\rm D}_A^{(z_{2})}W_S[C_{z_1z_2}] = -W_S[C_{z_1z_2}]{\cal A}_A(z_2)
                 \,.\label{eq.sderivative}
\eeqn
By virtue of the condition ${\cal A}_{\dot{\al}}=0$, $W_S[C_{z_1z_2}]$ 
satisfies 
\beqn
{\rm D}_{\dot{\al}}W_S[C_{z_1z_2}]=0\,.             \label{eq.chirality}
\eeqn

(D) We can act the area derivative in superspace on $W_S$ in a similar way to
   the bosonic case.    We find  
\beqn
& &\2{\Sig^{NM}(z_1)}W_S[C_{z_1z_2}]
   ={\cal F}_{NM}(z_1)W_S[C_{z_1z_2}] \,,\non\\
& &\de\Sig^{NM}=-\frac{1}{2} \delta s \delta t 
     \{\frac{\partial u^M}{\partial s}
       \frac{\partial u^N}{\partial t} 
       -(-)^{|M||N|} \frac{\partial u^N}{\partial s}
        \frac{\partial u^M}{\partial t} \}   \,.              \label{eq.sareaE}
\eeqn
 From the definition for the field strength ${\cal F}$, 
\beqn
{\cal F}=\1{2}dz^Mdz^N{\cal F}_{NM} 
        =  \1{2}e^B e^A {\cal F}_{AB}\,,
\eeqn
we may relate the components of ${\cal F}_{AB}$ with those of ${\cal F}_{NM}$
\beqn
{\cal F}_{AB}=-(-)^{|B||M|}e_A\,^M e_B\,^N {\cal F}_{NM}\,, 
\eeqn
where the matrix $e_A\,^M$ is defined by (\ref{matrix.117}). Defining the 
area derivative $\2{\Sig^{AB}}$ having flat indices by
\beqn
\2{\Sig^{AB}}\equiv-(-)^{|B||M|}e_A\,^M e_B\,^N \2{\Sig^{NM}}\,,
\eeqn 
we rewrite (\ref{eq.sareaE}) with respect to the flat indices:
\beqn
\2{\Sig^{AB}(z_1)}W_S[C_{z_1z_2}]
&=&{\cal F}_{AB}(z_1)W_S[C_{z_1z_2}]    \,.                   \label{eq.sareaF}
\eeqn
As we impose the flatness condition (\ref{eq.129}) on the one-form ${\cal A}$,
  we have
\beqn
\2{\Sig^{\al\be}}W_S=\2{\Sig^{\dal\dbe}}W_S=\2{\Sig^{\al\dal}}W_S=0 \,.
                                                           \label{eq.flatness}
\eeqn

(E) The supergauge transformations of the components ${\cal A}_M$ of
 ${\cal A}$ are
\beqn
{\cal A}_M\longrightarrow{\cal A}^{\prime}_M=-X^{-1}\pal_MX
                                       +X^{-1}{\cal A}_MX\,, \non
\eeqn
where $X$ is an element of the super gauge group. The $\phi(z)$ in 
(\ref{eq.sdiff}) transforms as
\beqn
\phi(z)\longrightarrow\phi^{\prime}(z)=X^{-1}\phi(z)\,.
\eeqn
 From these transformation laws, we obtain
\beqn
W_S[C_{z_1z_2}]\longrightarrow W_S^{\prime}[C_{z_1z_2}]
=X^{-1}(z_1)W_S[C_{z_1z_2}]X(z_2)     \,. \label{eq.trsW}
\eeqn

 We now take the special solution to the flatness condition (\ref{eq.129})
and look at the components of ${\cal A}$ having flat indices given by 
(\ref{eq.Ax}). The supergauge transformations  which keep  this condition are
 more restricted  than the most general supergauge transformations as
 is explained in Appendix C.
 The transformation laws of $W_S$
\beqn
W_S[C_{z_1z_2}]\longrightarrow W_S^{\prime}[C_{z_1z_2}]
={\rm e}^{-\Lambda(z_1)}W_S[C_{z_1z_2}] {\rm e}^{\Lambda(z_2)}    \,,
\eeqn
where $\Lambda (z)$ is a chiral superfield. The above 
transformation is consistent with (\ref{eq.chirality}) as
\beqn
{\rm D}_{\dot{\al}}W_S^{\prime}[C_{z_1z_2}]
&=&{\rm e}^{-\Lambda(z_1)}{\rm D}_{\dot{\al}}W_S[C_{z_1z_2}] 
{\rm e}^{\Lambda(z_2)}  \non\\
&=&0\,.\non
\eeqn 
It is clear that, in the case that the curve $C_{z_1z_2}$ is a closed loop
 $ C_{zz}$ in superspace, $trW_S[C_{zz}]$ is invariant under 
${\rm e}^{\rm V}\to {\rm e}^{{\rm V}^{\prime}}={\rm e}^{\Lambda^{\dagger}}
{\rm e}^{\rm V}{\rm e}^{\Lambda}$.

\section{The Schwinger-Dyson equation in the abelian Case}

  As is well-known, the supersymmetric 
pure Yang-Mills Lagrangian reads as
\beqn
{\cal L}_{SYM}
&=&\1{8g^2}tr({\cal W}{\cal W})\mid_{\mbox{$\te\te$ component}}   \non\\
&=&\1{8g^2}\int d^2\te tr({\cal W}{\cal W})    
 =\1{8g^2}\int d^2\te d^2\bar{\te}\,\de(\bar{\te})tr({\cal W}{\cal W})\,.
                                                             \label{eq.SYMlag2}
\eeqn 
Furthermore
\beqn
{\cal L}_{SYM}
&=&\1{8g^2}\int d^2\te d^2\bar{\te}\,\de(\bar{\te}) 
  tr\{-\1{4}\bar{\rm D}\bar{\rm D}{\rm e}^{-{\rm V}}{\rm D}^{\al}
  {\rm e}^{\rm V}{\cal W}_{\al} \}           \non\\
&=&-\1{32g^2}\int d^2\te d^2\bar{\te}\,
  (\bar{\rm D}\bar{\rm D}\de(\bar{\te}))
  tr\{{\rm e}^{-{\rm V}}{\rm D}^{\al}{\rm e}^{\rm V}{\cal W}_{\al} \} \non\\ 
&=&-\1{8g^2}\int d^2\te d^2\bar{\te}\,tr\{{\cal A}^{\al}{\cal W}_{\al} \}  
                                                        \,,  \label{eq.SYMlag3}
\eeqn 
where we have used $\bar{\rm D}_{\dot{\al}}{\cal W}_{\al}=0$ and omitted 
spacetime total derivatives.

Let us define the functional derivative of the vector superfield ${\rm V}$ 
by
\beqn
\2{{\rm V}(z)}
&=&\te^2\bar{\te}^2 \{\2{{\rm C}(x)}-\1{2}\Box\2{{\rm D}(x)} \}   \non\\
& &+\frac{2}{i}\te^2\bar{\te}_{\dot{\be}}\{\2{\bar{\chi}^{\dot{\be}}(x)}
   -\frac{i}{2}\pal_a\2{\lambda^{\al}(x)}\bar{\sig}^{a\dot{\be}\al} \}
   -\frac{2}{i}\bar{\te}^2\te^{\be}\{\2{\chi^{\be}(x)}
   -\frac{i}{2}\pal_a\2{\bar{\lambda}_{\dot{\al}}(x)}\sig^a_{\be\dot{\al}} \}
                                                                  \non\\
& &+\frac{2}{i}\bar{\te}^2\2{({\rm M}(x)+i{\rm N}(x))}
   -\frac{2}{i}\te^2\2{({\rm M}(x)-i{\rm N}(x))} \non\\
& &+2\te\sig^a\bar{\te}\2{v_a(x)}  
   -\frac{2}{i}\bar{\te}_{\dot{\al}}\2{\bar{\lambda}_{\dot{\al}}(x)}
   +\frac{2}{i}\te^{\al}\2{\lam^{\al}(x)}+2\2{{\rm D}(x)} \;.
\eeqn
In terms of $(y,\te,\bar{\te})$,
\beqn
\2{{\rm V}(z)}
&=&\te^2\bar{\te}^2 \2{{\rm C}(y)}
   -2i\te^2\bar{\te}_{\dot{\be}}\2{\bar{\chi}_{\dot{\be}}(y)}
   +2i\bar{\te}^2\te^{\be}\2{\chi^{\be}(y)}
   -2i\bar{\te}^2\2{({\rm M}(y)+i{\rm N}(y))}  \non\\
& &+2i\te^2\2{({\rm M}(y)-i{\rm N}(y))}
   +2\te\sig^a\bar{\te}\2{v_a(y)}+i\te^2\bar{\te}^2\pal_a\2{v_a(y)}
   +2i\bar{\te}_{\dot{\al}}\2{\bar{\lambda}_{\dot{\al}}(y)} \non\\
& &-2i\te^{\al}\{\2{\lam^{\al}(y)}
   -2i\te\sig^a\bte\pal_a\2{\lam^{\al}(y)} \}\non\\
& &+2\{\2{{\rm D}(y)}-i\te\sig^a\bte\pal_a\2{{\rm D}(y)}\}\,.\label{eq.d/dV}
\eeqn
These derivatives satisfy
\beqn
\2{{\rm V}(z)}{\rm V}(z^{\prime})
&=&\de^4(x-x^{\prime})\de(\te-\te^{\prime})\de(\bar{\te}-\bar{\te}^{\prime})
                                                                       \non\\
&=&\de^4(y-y^{\prime})\de(\te-\te^{\prime})\de(\bar{\te}-\bar{\te}^{\prime})
         =\de(z-z ^{\prime}) \,, \label{eq.dVg}
\eeqn
where $y^a=x^a+i\te\sig^a\bar{\te}$, 
$y^{\prime a}=x^{\prime a}+i\eta\sig^a\bar{\eta}$.


We are now ready to obtain an equation for the abelian case.  The field
${\rm V}$ includes the usual gauge field $v_a$ and their superpartners, 
gauginos $\lam$, $\bar{\lam}$ linearly.  As in
(\ref{eq.start})   we start out from
\beqn
0=\int[d{\rm V}]\2{{\rm V}(\zp)}\{{\rm e}^{i{\rm S}_{SA}}W_S[C] \}\,,
                                                          \label{eq.SABS-D}\\
{\rm S}_{SA}=\1{8g^2}\int d^4x d^2\te d^2\bar{\te}{\cal A}^{\al}{\cal W}_{\al}
                                                      \,,\label{eq.SABaction}\\
W_S[C]=P\exp\oint_C{\cal A}= \exp\oint_C{\cal A} \,.   \label{eq.SABWilson}
\eeqn
Note that, in  the abelian case, ${\cal A}$ as well as ${\cal W}_{\al}$
 is linear in the vector superfield ${\rm V}$,
\beqn
{\cal A}_{\al}&=&-{\rm D}_{\al}{\rm V}\,,\quad{\cal A}_{\dot{\al}}=0  \non\\
{\cal A}_a&=&\frac{i}{4}\bar{\rm D}\bar{\sig}^a{\rm D}{\rm V}\,,\non\\
{\cal W}_{\al}&=&-\1{4}\bar{\rm D}\bar{\rm D}{\rm D}_{\al}{\rm V}\;.
                                                    \label{eq.linearAW}
\eeqn
 We find
\beqn
\2{{\rm V}(\zp)}{\rm e}^{i{\rm S}_{SA}}
&=&i\1{32g^2}{\rm e}^{i{\rm S}_{SA}}\2{{\rm V}(\zp)}
\int d^4x d^2\te d^2\bar{\te}{\rm D}^{\al}{\rm V}\bar{\rm D}\bar{\rm D}
{\rm D}_{\al}{\rm V}                            \non\\
&=&i\1{16g^2}{\rm e}^{i{\rm S}_{SA}}\int d^4x d^2\te d^2\bar{\te}
({\rm D}^{\al}\de(z-z^{\prime}))\bar{\rm D}\bar{\rm D}
{\rm D}_{\al}{\rm V}   \non\\
&=&-i\1{4g^2}{\rm D}^{\al}{\cal W}_{\al}(\zp){\rm e}^{i{\rm S}_{SA}} \,.
                                                          \label{eq.eq.of.mot.}
\eeqn
The components of ${\rm D}^{\al}{\cal W}_{\al}={\rm D}{\cal W}=0$ correspond
to field equations for the fields 
$(\lam^{\al},\lam_{\dot{\al}},v_a,{\rm D})$ in (\ref{eq.vector}), 
\beqn
{\rm D}{\cal W}&=&-2D(y)+2\te\sig^a\pal_a\bar{\lam}(y)
                  -2\bar{\te}\bar{\sig}^a\pal_a\lam(y)       \non\\
               & &-2\te\sig_a\bar\te\pal_b v^{ba}(y)
                  +2i\te\te\bar{\te}\Box\bar{\lam}(y)           
                  +2i\te\sig^a\bte\pal_a{\rm D}(y)            \non\\
               &=&0                                              \non\\
&\Longleftrightarrow& \left\{
                 \begin{array}{r}               
                      {\rm D}=0\,\,,  \\
                 \sig^a\pal_a\bar{\lam}=0\,\,,\bar{\sig}^a\pal_a\lam=0\,\,,\\
                 \pal_a v^{ab}=0\,\,.     
                 \end{array}\right.                     \label{eq.eq.of.mot.DW}
\eeqn
From the relation ${\cal F}_{a\dal}=i\sig_{a\al\dal}{\cal W}^{\dal}/2$, 
we may rewrite (\ref{eq.eq.of.mot.}) as
\beqn
\frac{\de}{\de{\rm V}(\zp)}{\rm e}^{i{\rm S}_{SA}}
=\frac{1}{8g^2}\ep_{\al\be}\bar{\sig}^{a\dot{\al}\be}\{{\rm D}^{\al}
                     {\cal F}_{a\dot{\al}}(\zp)\}{\rm e}^{i{\rm S}_{SA}}\,.
\eeqn
Remember that the field strength ${\cal F}_{a\dot{\al}}$ is obtained 
by the area derivative on $W_S[C]$. Using this fact and 
(\ref{eq.eq.of.mot.}), we find 
\beqn
\{\frac{\de}{\de{\rm V}(\zp)}{\rm e}^{i{\rm S}_{SA}}\} W_S[C]
=\frac{1}{8g^2}\ep_{\al\be}\bar{\sig}^{a\dot{\al}\be}\{{\rm D}^{\al}
        \frac{\de}{\de\Sigma^{a\dot{\al}}(\zp)} W_S[C]\}
                                 {\rm e}^{i{\rm S}_{SA}}\,.   \label{eq.lsS-D}
\eeqn
On the other hand,
\beqn
\frac{\de}{\de{\rm V}(\zp)}W_S[C]
&=&\oint_C\{\frac{\de}{\de{\rm V}(\zp)}{\cal A}(z(t))\}W_S[C] \non\\
&=&W_S[C]\oint_C\{\frac{i}{4}(dx^a-id\eta\sig^a\bar{\eta}  
+i\eta\sig^ad\bar{\eta})\bar{\rm D}_{\eta}\bar{\sig}_a{\rm D}_{\eta}  \non\\
& &\hspace{1cm}-d\eta{\rm D}_{\eta}\}\de(z-\zp)  \,.    \label{eq.rsS-D}
\eeqn
In terms of $(y,\te,\bar{\te})$,
\beqn
\frac{\de}{\de{\rm V}(\zp)}W_S[C]
&=&W_S[C]\oint_C\{\frac{i}{4}(dy^a-2id\eta\sig^a\bar{\eta})
 \bar{\rm D}_{\eta}\bar{\sig}_a{\rm D}_{\eta}  \non\\
& &\hspace{1cm}-d\eta{\rm D}_{\eta}\}\de(z-\zp)\,,
                                                              \label{eq.rsS-Dx}
\eeqn
where $z=(x,\eta,\bet)$, $z^{\prime}=(x^{\prime},\te,\bar{\te})$. We denote by
${\rm D}_{\eta}$, $\bar{\rm D}_{\eta}$ the superderivatives:
\beqn
{\rm D}_{\eta\al}
&=&\frac{\pal}{\pal\eta^{\al}}+i(\sig^a\bar{\eta})_{\al}
                                   \frac{\pal}{\pal x^{\prime a}} \,,  \non\\
\bar{\rm D}_{\eta\dot{\al}}
&=&-\frac{\pal}{\pal\bar{\eta}^{\al}}-i(\eta\sig^a)_{\dot{\al}}
                                    \frac{\pal}{\pal x^{\prime a}} \,.
\eeqn
Considering both (\ref{eq.lsS-D}) and (\ref{eq.rsS-D}), we obtain  the  
supersymmetric Schwinger-Dyson equation in the abeian case:
\beqn
\frac{1}{8g^2}\ep_{\al\be}\bar{\sig}^{a\dot{\al}\be}{\rm D}^{\al}
        \frac{\de}{\de\Sigma^{a\dot{\al}}(\zp)} {\bf W}_S[C]            
={\bf W}_S[C]\oint_C{\cal D}_f^{\al}{\rm D}_{\eta\al}\de(z-\zp)\,,
                                                                      \non\\
{\bf W}_S[C] \equiv \lan W_S[C] \ran 
   = \int[d{\rm V}]{\rm e}^{i{\rm S}_{SA}}W_S[C]\,, \label{eq.SABM-M}
\eeqn
where we have introduced the derivative ${\cal D}_f^{\al}$ by
\beqn
{\cal D}_f^{\al}
&=&-\frac{i}{4}(dx^a-id\eta\sig^a\bar{\eta}+i\eta\sig^a d\bar{\eta})
       \bar\sig_a^{\dal\al}\bar{\rm D}_{\eta\dot{\al}}   \non\\
& &+d\eta^{\al}   \non\\
&=&-\im{4}(dy^a-2id\eta\sig^a\bet)\bsi^{\dal\al}_a\bar{\rm D}_{\eta\dal}
                                        +d\eta^{\al}  \non\\
&=&-e^a\frac{i}{4}\bar\sig_a^{\dal\al}\bar{\rm D}_{\eta\dot{\al}}+e^{\al}
                                                      \,.\label{eq.Df}
\eeqn


\section{The Schwinger-Dyson Equation in the nonabelian case}

 Let us consider the
supersymmetric Schwinger-Dyson equation for nonabelian gauge group $U(N_c)$. 
In the previous section both the one-form ${\cal A}$ and  the chiral 
superfield ${\cal W}_{\al}$  were linear in the vector superfield ${\rm V}$. 
In this case,  it is natural to start with (\ref{eq.SABS-D}). 
In the nonabelian case, both ${\cal A}$ and ${\cal W}_{\al}$ are highly
 non-linear  in the fundamental superfield ${\rm V}$.
 The functional derivatives of ${\rm V}$ acting on the super Wilson-loop and
the super Yang-Mills action become very 
complicated. It is technically difficult to obtain this way
  the nonabelian counterpart of eq.~(\ref{eq.SABM-M}) in the last section
 which is composed  only
 of the super Wilson-loops and  the geometric operations.
  We must find another way to obtain the equation.

Let us pay attention to the left hand side of equation 
(\ref{eq.nontrivial}):
\beqn
i\1{4g^2}\lan tr\{D_b v^{ba}(x)W[C_{xx}] \}\ran \,.\label{eq.eq.W}
\eeqn
This includes equation of motion for the Yang-Mills field $v_a$ i.e.
$D_b v^{ba}$.  It 
is obtained by hitting the area derivative and  subsequently the ordinary derivative on $\lan W_S[C] \ran$. (See Appendix B).

In the previous section, the term ${\rm D}^{\al}{\cal W}_{\al}$
 appeared in (\ref{eq.eq.of.mot.}). As in (\ref{eq.eq.of.mot.}), each
 component of ${\rm D}^{\al}{\cal W}_{\al}$
 is composed of equations of motion for super abelian theory, and this is 
obtained by
\beqn
\1{8g^2}\ep_{\al\be}\bar{\sig}^{a\dot{\al}\be}{\rm D}^{\al}
   \2{\Sig^{a\dot{\al}}(z)}\lan W_S[C] \ran \,.
\eeqn 
 We infer that the final form of the supersymmetric 
equation should include
\beqn
\1{8g^2}\ep_{\al\be}\bar{\sig}^{a\dot{\al}\be}{\rm D}^{\al}
   \2{\Sig^{a\dot{\al}}(z)}\lan tr W_S[C] \ran   \label{eq.start of SYMS-D}
\eeqn 
on the lefthand side.
We choose (\ref{eq.start of SYMS-D}) our starting point for  the
nonabelian case.
Using (\ref{eq.sareaF}) and (\ref{eq.sderivative}), we see that the above 
equation (\ref{eq.start of SYMS-D}) becomes
\beqn
& &\1{8g^2}\ep_{\al\be}\bar{\sig}^{a\dot{\al}\be}{\rm D}^{\al}
   \2{\Sig^{a\dot{\al}}(z)}\lan tr W_S[C] \ran     \non\\
&=&\1{8g^2}\ep_{\al\be}\bar{\sig}^{a\dot{\al}\be}{\rm D}^{\al}
   \lan tr\{{\cal F}_{a\dot{\al}}(z)W_S[C_{zz}] \}\ran   \non\\
&=&\1{8g^2}\ep_{\al\be}\bar{\sig}^{a\dot{\al}\be}
   \lan tr({\cal D}^{\al}{\cal F}_{a\dot{\al}}(z)W_S[C_{zz}])\ran  \non\\
&=&-\frac{i}{4g^2}\lan tr\{{\cal D}{\cal W}(z) W_S[C_{zz}] \}\ran\,,
                                                    \label{eq.start of SYM}
\eeqn
\beqn
  {\rm where} \;\;\;\;
{\cal D}{\cal W} &\equiv& {\rm D}^{\al}{\cal W}_{\al}
                     -\{{\cal A}^{\al},{\cal W}_{\al} \}\,,
                                                        \label{eq.eq.of.SYM} \\
{\cal F}_{a\dal} &=&  \im{2}\sig_{a\al\dal}{\cal W}^{\al}\,.\label{eq.W and F}
\eeqn
  The expression ${\cal D}{\cal W}$ is the 
nonabelian extension of ${\rm D}^{\al}{\cal W}_{\al}$. The components are
equations of motion of the super Yang-Mills theory.
Under the transformation of ${\rm V}$,  (\ref{eq.sgaugetr}),
\beqn
{\rm e}^{\rm V}\longrightarrow{\rm e}^{{\rm V}^{\prime}}
={\rm e}^{{\Lambda}^{\dagger}}{\rm e}^{\rm V}{\rm e}^{\Lambda}  \,,
                                                         \label{eq.sgaugetr2}
\eeqn
 the superfield (\ref{eq.eq.of.SYM}) 
transforms covariantly,
\beqn
{\cal D}{\cal W}\longrightarrow({\cal D}{\cal W})^{\prime}
 = {\rm e}^{-\Lambda}{\cal D}{\cal W}{\rm e}^{\Lambda} \label{eq.sgaugetrDW}
\eeqn
  The expression
 (\ref{eq.start of SYM}) is  locally super-gauge invariant. 
 In order to make further manipulations manageable,
  we employ  the Wess-Zumino gauge condition, 
which eliminates the fields $({\rm C},\chi,\bar{\chi},{\rm M},{\rm N})$.
 We will mainly use the coordinates 
$(y^a=x^a+i\te\sig^a\bar{\te}, \te, \bar{\te})$ as
 the gauge covariance becomes clearer with these by the following reason.
 The chiral superfield 
$\Lambda$  is  the parameter of the restricted supergauge transformation. The
imaginary part $u$ of the lowest component of $\Lambda$ corresponds to
  the usual local 
gauge parameter depending on the spacetime coordinates.
  In  the Wess-Zumino gauge, only $u$ of 
$\Lambda$ survives as gauge parameters, in terms of $(y,\te,\bar{\te})$. The 
transformations (\ref{eq.135}) and (\ref{eq.126}) become
\beqn
{\cal A}_A&\longrightarrow&
{\cal A}^{\prime}_A=-{\rm e}^{iu(y)}{\rm D}_A{\rm e}^{-iu(y)}
                    +{\rm e}^{iu(y)}{\cal A}_A{\rm e}^{-iu(y)}\,,\non\\
{\cal W}_{\al}&\longrightarrow&
{\cal W}^{\prime}_{\al}={\rm e}^{iu(y)}{\cal W}_{\al}{\rm e}^{-iu(y)}\,,
                                                      \label{eq.usualgauge}
\eeqn

We first calculate  the one-form ${\cal A}$ and the chiral superfield 
${\cal W}_{\al}$ in the Wess-Zumino gauge, where
\beqn
{\rm V}
&=&-\te\sig^a\bte v_a(x)+i\te^2\bte\bla(x)-i\bte^2\te\lam(x)  \non\\
& &+\1{2}\te^2\bte^2{\rm D}(x)                            \non\\
&=&-\te\sig^a\bte v_a(y)+i\te^2\bte\bla(y)-i\bte^2\te\lam(y)  \non\\
& &+\1{2}\te^2\bte^2{\rm D}(y)-\im{2}\te^2\bte^2\pal_a v^a(y)\,,
\eeqn
\beqn
{\rm e}^{\rm V}
&=&1+{\rm V}+\1{2}{\rm V}^2  \non\\
&=& 1-\te\sig^a\bte v_a(y)+i\te^2\bte\bla(y)-i\bte^2\te\lam(y) \non\\
& &+\1{2}\te^2\bte^2{\rm D}(y)-\1{4}\te^2\bte^2 v^2(y)
-\im{2}\te^2\bte^2\pal_a v^a(y)\,, \non\\                    \label{eq.e^V}\\
{\rm e}^{-\rm V}
&=&1-{\rm V}+\1{2}{\rm V}^2  \non\\
&=&1+\te\sig^a\bte v_a(y)-i\te^2\bte\bla(y)+i\bte^2\te\lam(y) \non\\
& &-\1{2}\te^2\bte^2{\rm D}(y)+\1{4}\te^2\bte^2 v^2(y)
-\im{2}\te^2\bte^2\pal_a v^a(y) \,.\non\\                   \label{eq.e^(-V)}
\eeqn
Using the expressions of ${\rm D}_{\al}$, $\bar{\rm D}_{\dal}$ in terms of 
$(y,\te,\bte)$, we obtain ${\cal A}_{\al}$ and
${\cal A}_a$:
\beqn
\xi^{\al}{\cal A}_{\al}
&=&-{\rm e}^{-\rm V}\xi^{\al}{\rm D}_{\al}{\rm e}^{\rm V}       \non\\
&=&\xi\sig^a\bar{\te}v_a(y)      
   -2i\xi\te\bar{\te}\bar{\lam}(y)+i\bar{\te}^2\xi\lam(y) \non\\
& &-\xi\te\bar{\te}^2{\rm D}(y)+i\bar{\te}^2\xi\sig^{ab}\te v_{ab}(y)\non\\
& &-\te^2\bar{\te}^2\xi\sig^a D_a\bar{\lam}(y)\,,            \label{eq.A_(al)}\\
{\cal A}_a
&=&\frac{i}{4}\bsi^{\dal\al}_a \bar{\rm D}_{\dal}{\rm e}^{-\rm V}
                {\rm D}_{\al}{\rm e}^{\rm V}                      \non\\
&=&-\im{2}v_a(y) -\1{2}\bla(y)\bsi_a\te+\1{2}\bte\bsi_a\lam(y)  \non\\
& &+\im{2}\bte\bsi_a\te{\rm D}(y)+\1{2}\bte\bsi_a\sig^{bc}\te v_{bc}(y)\non\\
& &+\im{2}\te^2\bte\bsi_a\sig^b D_b\bla(y)\,,                   \label{eq.A_a}
\eeqn
where we have defined $v_{ab}(y)$ and $D_a\lam(y)$ respectively by 
\beqn
v_{ab}(y)&=&\pal_a v_b(y)-\pal_b v_a(y)+\im{2}[v_a(y),v_b(y)] \,,\non\\
D_a\lam(y)&=&\pal_a\lam(y)+\im{2}[v_a(y),\lam(y)]\,.\non
\eeqn
 From the gauge transformation of ${\cal A}_a$ given by (\ref{eq.usualgauge}) 
and the explicit expression (\ref{eq.A_a}), we find
\beqn
v_a(y)&\longrightarrow&v^{\prime}_a(y)=-2iU(y)\pal_a U^{-1}(y)
                                       +U(y)v_a(y)U^{-1}(y)\,,\non\\
\lam(y)&\longrightarrow&\lam^{\prime}(y)=U(y)\lam(y)U^{-1}(y)\,,\non\\
\bla(y)&\longrightarrow&\bla^{\prime}(y)=U(y)\bla(y)U^{-1}(y)\,,\non\\
{\rm D}(y)&\longrightarrow&
                     {\rm D}^{\prime}(y)=U(y){\rm D}(y)U^{-1}(y)\,,\non
\eeqn
where $U(y)$ denotes ${\rm e}^{iu(y)}$.

The chiral superfield ${\cal W}_{\al}$ takes the form
\beqn
\xi^{\al}{\cal W}_{\al}
&=&-\1{4}\bar{\rm D}\bar{\rm D}{\rm e}^{-\rm V}\xi^{\al}{\rm D}_{\al}
                                              {\rm e}^{\rm V}  \non\\
&=&-i\xi\lam(y)-i\xi\sig^{ab}\te v_{ab}(y)+\xi\te{\rm D}(y)\non\\
& &+\te^2\xi\sig^a D_a\bla(y)\,.     \label{ex.W_(al)} 
\eeqn
Note that
$\bar{\rm D}_{\dal}{\cal W}_{\al}=0$ is trivially satisfied.
 
Inserting (\ref{eq.A_(al)}) into the one-form ${\cal A}$ given by
 (\ref{eq.Ay}), we obtain 
\beqn
{\cal A}
&=&dy^a\{-\im{2}v_a(y)-\1{2}\bla(y)\bsi_a\eta+\1{2}\bet\bsi_a\lam(y) \non\\
& &+\im{2}\bet\sig_a\eta{\rm D}(y)+\1{2}\bet\bsi_a\sig^{bc}\eta v_{bc}(y)
                       +\im{2}\eta^2\bet\bsi_a\sig^b D_b\bla(y) \}    \non\\
& &+\bet^2\{-id\eta\lam(y)+d\eta\eta{\rm D}(y)
                           -id\eta\sig^{ab}\eta v_{ab}  \non\\
& &+\eta^2d\eta\sig^a D_a\bla (y) \} \,.  \label{eq.1-form A}             
\eeqn

The super Yang -Mills action is derived from (\ref{eq.SYMlag2}),
\beqn
{\rm S}_{SYM}
&=&\1{4g^2}\int d^4y \,tr\{-i\lam(y)\sig^a D_a\bla(y)+\1{2}{\rm D}^2(y)\non\\
& &\hspace*{2cm}-\1{4}v_{ab}(y)v^{ab}(y)
      -\im{8}\ep^{abcd}v_{ab}(y)v_{cd}(y) \}    
\eeqn
Hereafter, we denote a closed loop $C$ on the superspace by
\beqn
z=z(t)=(y(t),\eta(t),\bet(t)), \;\;\;0\le t\le 1\,.
\eeqn 
A point on the loop $C$ is denoted by $z^{\prime}=(y^{\prime},\te,\bte)$.
Let us take functional derivatives with respect to 
$v_a(\yp),\,\lam(\yp),\,\bla(\yp)$ and ${\rm D}(\yp)$ acting the action
 ${\rm S}_{SYM}$ and obtain equations of motion for these fields:
\beqn
T^r\2{v^{(r)}_a(\yp)}{\rm S}_{SYM}
&=&\1{4g^2}\left(D_b v^{ba}(\yp)
-\1{2}\bsi^{a\dal\al}\{\bla_{\dal}(\yp),\lam_{\al}(\yp)\}\right)\,,\non\\
T^r\2{\lam^{(r)\al}(\yp)}{\rm S}_{SYM}
&=&-\im{4g^2}(\sig^a D_a\bla(\yp))_{\al}\,,\non\\
T^r\2{\bla^{(r)}_{\dal}(\yp)}{\rm S}_{SYM}
&=&-\im{4g^2}(\bsi^a D_a\lam(\yp))^{\dal}\,,\non\\
T^r\2{{\rm D}^{(r)}(\yp)}{\rm S}_{SYM}
&=&\1{4g^2}{\rm D}(\yp)\,,                        \label{eq.eq.of.mot.SYM}
\eeqn
where $T^r$ denotes the generators of the gauge group $U(N_c)$ and
 the summation over $r$ is implied.
 The expression${\cal DW}$  in (\ref{eq.start of SYMS-D}) is given in
 the Wess-Zumino gauge by
\beqn
{\cal D}{\cal W}
&=&{\rm D}^{\al}{\cal W}_{\al}-\{{\cal A}^{\al},{\cal W}_{\al} \}\non\\
&=&-2{\rm D}(y)-2\bte\bsi^a D_a\lam(y)+2\te\sig^a D_a\bla(y)\non\\
& &-2\te\sig_b\bte[D_a v^{ab}(y)
      -\1{2}\bsi^{b\dal\al}\{\bla_{\dal}(y),\lam_{\al}(y) \}]\non\\
& &+2i\te\sig^a\bte D_a{\rm D}(y)-2i\te^2\bte\bsi^b\sig^a D_b D_a\bla(y)\non\\
& &-2i\te^2[{\rm D}(y),\bte\bla(y)]  \,.                    \label{eq.eprDW}
\eeqn
Using (\ref{eq.eq.of.mot.SYM}) and (\ref{eq.eprDW}), we can rewrite the 
right hand side of equation (\ref{eq.start of SYMS-D}) as
\beqn
& &-\im{4g^2}\lan tr\{{\cal D}{\cal W}(z^{\prime})
             W_S[C_{z^{\prime}z^{\prime}}]\}\ran \non\\
&=&-\im{4g^2}\int [dv_ad\lam^{\al}d\bla_{\dal}d{\rm D}]{\rm e}^{i{\rm S}_{SYM}}
   tr\{{\cal D}{\cal W}(z^{\prime})W_S[C_{z^{\prime}z^{\prime}}]\} \non\\
&=&2\int [dv_ad\lam^{\al}d\bla_{\dal}d{\rm D}]
   tr\{T^r\,W_S[C_{z^{\prime}z^{\prime}}]\}
    \2{{\rm V}^{(r)}_{\mbox{mod}}(z^{\prime})}{\rm e}^{i{\rm S}_{SYM}}
                                                    \label{eq.lhs.of.SS-D}
\eeqn
\beqn
\2{{\rm V}^{(r)}_{\mbox{mod}}(z^{\prime})}
&\equiv&\2{{\rm D}^{(r)}(y^{\prime})}
        +i\bte_{\dal}\2{\bla_{\dal}^{(r)}(y^{\prime})}
        -i\te^{\al}\2{\lam^{(r)\al}(y^{\prime})}  \non\\
& &+\te\sig_a\bte\2{v_a^{(r)}(y^{\prime})}
   -i\te\sig^a\bte D_a^{{\prime}(rs)}\2{{\rm D}^{(s)}(y^{\prime})}   \non\\
& &-\te^2(\bte\bsi^a)^{\al} D_a^{{\prime}(rs)}\2{\lam^{(s)\al}(y^{\prime})}
-i\te^2\bte\bla^{(rs)}(y^{\prime})\2{{\rm D}^{(s)}(y^{\prime})}\,,\non \\
& &                         \label{eq.d/dV_m}
\eeqn
and
\beqn
D_a^{{\prime}(rs)}
&\equiv&\de^{rs}\frac{\pal}{\pal y^{\prime a}}+\im{2}v^{(rs)}_a(y^{\prime})\,,
                                                                        \non\\
v^{(rs)}_a(y^{\prime})&\equiv&if^{rts}v^{(t)}_a(y^{\prime})\,,\non\\
\bla^{(rs)}(y^{\prime})&\equiv&if^{rts}\bla^{(t)}(y^{\prime})\,.
\eeqn
Note that $\2{{\rm V}^{(r)}_{\rm mod}}$ 
does not contain the derivatives with respect to the fields 
$({\rm C},\chi,\bar{\chi},{\rm M},{\rm N})$, while $\2{{\rm V}}$ defined by 
equation (\ref{eq.d/dV}) does. The covariant derivative $D_a^{(rs)}$ and fields
 $\bla^{(rs)}$ appear in the expression of $\2{{\rm V}^{(r)}_{\rm mod}}$. 
This is
because, even in the Wess-Zumino gauge, neither 
${\cal A}$ nor ${\cal W}_{\al}$  is linear in ${\rm V}$ in  the
nonabelian case.
 
The infinitesimal change of the one-form
${\cal A}$ yields 
\beqn
\de W_S[C_{z_1z_2}]=\int_0^1 W_S[C_{z_1z(t)}]\de{\cal A}(z(t))
                             W_S[C_{z(t)z_2}]\,,\label{eq.variationW_S}
\eeqn 
where $\int_0^1$ denotes the integration over $t$ which parameterizes the 
curve $C_{z_1z_2}$ and the line element $\frac{dz^M}{dt}$ is included in the 
expression $\de{\cal A}(z(t))$.

Let us carry out the partial integration of the functional variables 
 $(v_a,\lam,\bla,{\rm D})$ in (\ref{eq.lhs.of.SS-D}).
 The functional derivative
$\2{V_{\rm mod}}$ acts on the Wilson-loop $W_S[C_{\zp\zp}]$.
 Using 
 (\ref{eq.variationW_S}), we find
\beqn
& &2\int [dv_ad\lam^{\al}d\bla_{\dal}d{\rm D}]\,
   tr\{T^r\,W_S[C_{z^{\prime}z^{\prime}}]\}
    \2{{\rm V}^{(r)}_{\mbox{mod}}(z^{\prime})}{\rm e}^{i{\rm S}_{SYM}} \non\\
&=&-2\int [dv_ad\lam^{\al}d\bla_{\dal}d{\rm D}]{\rm e}^{i{\rm S}_{SYM}} \non\\
& &\hspace*{1cm}\times\oint \,tr\{T^r\,W_S[C_{z^{\prime}z}]
        \frac{\de{\cal A}(z)}{\de{\rm V}^{(r)}_{\mbox{mod}}(z^{\prime})}
                               W_S[C_{zz^{\prime}}]\}\,,\label{eq.rhs.of.SS-D}
\eeqn
where $\oint$ denotes the integration over $z$ along the closed loop 
$C_{z^{\prime}z^{\prime}}=C$ and the line element $dz^M$ is included in 
${\cal A}(z)$ as in (\ref{eq.variationW_S}).

After some calculation from (\ref{eq.1-form A}) and (\ref{eq.d/dV_m}), 
we obtain
\beqn
& &\2{{\rm V}^{(r)}_{\mbox{mod}}(z^{\prime})}{\cal A}(z)
={\cal K}_y+{\cal K}_{\eta}\,,\label{eq.K_y+K_(eta)}\\
{\cal K}_y
&\equiv&T^s\{-\im{2}\de^{rs}(\eta-\te)\sig^a(\bet-\bte) \non\\
& &-\1{2}dy^a\bte\bsi^b\sig_a\bet \eta^2 \tilde{D}_b^{(sr)}
-dy^b\bet\bsi_b\sig^{ca}\eta\bte\bsi_a\te\tilde{D}_c^{(sr)} \non\\
& &-\1{2}dy^b\eta\sig_b\bet\bte\bsi^a\te\tilde{D}_a^{(sr)}
-\1{2}dy^b\te^2\bte\bsi^a\sig_b\bet\tilde{D}_a^{(sr)}\}\de^4(y-y^{\prime})
                                                         \,,\label{eq.K_y}\\
{\cal K}_{\eta}
&\equiv&T^s\bet^2\{\de^{sr}(\eta d\eta-\te d\eta) \non\\
& &+i\eta^2d\eta\sig^a\bte \tilde{D}_a^{(sr)}
-2i\te\sig_a\bte d\eta\sig^{ba}\eta\tilde{D}_b^{(sr)} \non\\
& &+id\eta\eta\te\sig^a\bte\tilde{D}_a^{(sr)} 
+i\te^2d\eta\sig^a\bte\tilde{D}_a^{(sr)}\} \de^4(y-y^{\prime})
                                                           \label{eq.K_(eta)}  
\eeqn
\beqn
\tilde{v}_a\equiv v_a-i\bla\bsi_a\eta\,,\\
\tilde{D}_a^{(sr)}\equiv \de^{sr}\frac{\pal}{\pal y^a}
                            +\im{2}\tilde{v}_a^{(sr)}(y)
\eeqn
where the summation over $s$ is taken implicitly. 
Note that all derivatives in the above equations are  with respect to $y^a$; 
they act on the delta function $\de^4(y-\yp)$.

The right hand side of 
(\ref{eq.rhs.of.SS-D})  as it is contains the terms depending on the fields 
$\tilde{v}_a=v_a-i\bla\bsi_a\eta$ explicitly and is not represented by 
supersymmetric Wilson-loops alone. 
 Our remaining task in this section is to show that these fields are 
generated from geometrical operations acting on the super Wilson-loop
Using (\ref{eq.sderivative}), we find 
\beqn
& &\frac{\pal}{\pal y^a}T^r W_S[C_{\zp z}]T^r\de^4(y-\yp)W_S[C_{z \zp}]\non\\
&=&T^rW_S[C_{\zp z}]\{T^r(\pal_a\de^4(y-\yp))W_S[C_{z \zp}]
-[{\cal A}_a(z),T^r]\de^4(y-\yp) \}\non\\
&=&T^rW_S[C_{\zp z}]T^s\{{\cal D}_a^{(sr)}\de(y-\yp) \}W_S[C_{z \zp}] \,, \\
& & {\cal D}_a^{(sr)}\equiv \de^{sr}\frac{\pal}{\pal y^a}
         -{\cal A}_a^{(sr)}(z)\,.   \label{eq.d/dy W_S}
\eeqn 
Note that, in (\ref{eq.K_y+K_(eta)}), $\2{V^{(r)}_{\rm mod}(\zp)}{\cal A}(z)$ 
depends on the fields $v_a$ and $\bla_{\dal}$ only through their combination 
$\tilde{v}_a=v_a-i\bla\bsi_a\eta$. From (\ref{eq.A_a}), we can express 
${\cal A}_a(z)$ via $\tilde{v}_a$:
\beqn
{\cal A}_a(z)=-\im{2}\tilde{v}_a(y)
-\bet^{\dal}\bar{\rm D}_{\dal}{\cal A}_a(z)\,.
\eeqn
Furthermore, ${\cal A}_a(z)$ 
can be rewritten as
\beqn
{\cal A}_a(z)
&=&-\im{2}\tilde{v}_a(y)
-\im{4}\bet^{\dal}\bar{\rm D}_{\dal}\bsi_a^{\dbe\be}\bar{\rm D}_{\dbe}
    {\rm e}^{-\rm V}{\rm D}_{\be}{\rm e}^{\rm V}      \non\\
&=&-\im{2}\tilde{v}_a(y)
+\im{8}\bet^{\dal}\bsi_a^{\dbe\be}\ep_{\dal\dbe}\bar{\rm D}\bar{\rm D}
    {\rm e}^{-\rm V}{\rm D}_{\be}{\rm e}^{\rm V}      \non\\
&=&-\im{2}\tilde{v}_a(y)+\1{2}\bet\bsi_a{\cal W}(z)\,. \label{eq.A=v+W}
\eeqn
Here, in the last equality, we have used (\ref{eq.defW}).
 Using this expression, we can write 
(\ref{eq.K_y+K_(eta)}) in terms of ${\cal A}_a(z)$ and
 ${\cal W}_{\al}(z)$ without employing $\tilde{v}_a$,
\beqn
& &\2{{\rm V}^{(r)}_{\mbox{mod}}(z^{\prime})}{\cal A}(z) \non\\
&=&T^s\{-\im{2}\delta^{rs} dy^a(\eta-\te)\sig_a(\bet-\bte)
-\1{2}dy^a\de(\eta-\te)\bte\bsi^b\sig_a\bet{\cal D}_b^{(sr)} \non\\
& &+\im{4}\bet^2dy^a\bte\bsi_a{\cal W}^{(sr)} \non\\
& &+\bet^2\left(d\eta(\eta-\te) \delta^{rs}
+id\eta\sig^a\bte\de(\eta-\te){\cal D}_a^{(sr)}\right) \}
                                                    \de^4(y-y^{\prime})\,.
\eeqn
We see that all field dependent terms are expressed through
${\cal D}_a^{(sr)}$ and ${\cal W}_{\al}$. Note that, in equation 
(\ref{eq.A=v+W}), ${\cal W}_{\al}$ is multiplied by $\bet$, while  
${\cal K}_{\eta}$ defined by (\ref{eq.K_(eta)}) is multiplied by $\bet^2$. 
So ${\cal W}_{\al}$ originates from ${\cal K}_y$ only. 

Remember the relation between ${\cal W}_{\al}$ and 
${\cal F}_{a\dal}$ given by (\ref{eq.W and F}) and that 
${\cal F}_{a\dal}$ is obtained from area derivatives
 by acting on the super Wilson-loop. The right hand side of 
(\ref{eq.rhs.of.SS-D}) can, therefore, be expressed  with
the derivative $\pal_a$ and the area
 derivative of the super Wilson-loop alone. We conclude 
\beqn
& &\1{8g^2}\ep_{\al\be}\bar{\sig}^{a\dot{\al}\be}{\rm D}^{\al}
   \2{\Sig^{a\dot{\al}}(\zp)}\lan tr W_S[C] \ran  \non\\
&=&-2\oint \{-\im{2}dy^a(\eta-\te)\sig^a(\bet-\bte)
-\1{2}dy^a\de(\eta-\te)\bte\bsi^b\sig_a\bet\pal_b\non\\
& &\hspace*{0.7cm}+\bet^2\left(d\eta(\eta-\te)
+id\eta\sig^a\bte\de(\eta-\te)\pal_a\right) \}
\lan tr\,W_S[C_{z^{\prime}z}] \de^4(y-y^{\prime})
                tr\,W_S[C_{zz^{\prime}}]\ran \non\\
& &+2\oint\1{8}
\bet^2\de(\eta-\te)\de^4(y-y^{\prime})dy^a(\bsi^b\sig_a\bte)^{\dal}\non\\
& &\times \left\{\lan(\2{\Sig^{b\dal}(z)} tr\,W_S[C_{z^{\prime}z}])
                       tr\,W_S[C_{zz^{\prime}}]\ran
                    -\lan tr\,W_S[C_{z^{\prime}z}]
            (\2{\Sig^{b\dal}(z)}tr\, W_S[C_{zz^{\prime}}])\ran \right\}\non\\
& &      \label{eq.SS-D}
\eeqn
Here we have used the completeness relation obeyed by the generators
 $T^r$ of $U(N_c)$.

We have obtained the supersymmetric 
Schwinger-Dyson equation for  the nonabelian gauge groups
 (\ref{eq.SS-D}) in a closed form
 with respect to the super Wilson-loops.
Eq.(\ref{eq.SS-D}) is invariant under the ordinary  gauge transformation 
 given by (\ref{eq.usualgauge}).
 The invariance of the left hand side  is obvious while  that of the right hand
side is less obvious.
 It contains the delta function $\de^4(y-\yp)$.
 At the nonvanishing support,  the right hand side is invariant as well.

\section{Manifestly Supersymmetric Form of the Schwinger-Dyson Equation}

One may wonder whether (\ref{eq.SS-D}) is consistent with (\ref{eq.SABM-M})
 at $N_c=1$, namely, when we take the abelian
 reduction of the nonabelian equation.   We have
\beqn
& &\1{8g^2}\ep_{\al\be}\bsi^{a\dal\be}{\rm D}^{\al}\2{\Sig^{a\dal}(\zp)} 
                      \lan W_S[C] \ran   \non\\
&=&-2\lan W_S[C]\ran\oint\{-\im{2}dy^a(\eta-\te)\sig_a(\bet-\bte) \non\\
& &\hspace*{1cm}-\1{2}dy^a\de(\eta-\te)\bte\bsi^b\sig_a\bet\pal_b   \non\\
& &\hspace*{1cm}+\bet^2(d\eta(\eta-\te)+id\eta\sig^a\bte\de(\eta-\te)\pal_a)\}
                \de^4(y-\yp)\,. \label{eq.red of SS-D}
\eeqn
After some calculations, we see that the right hand side of 
(\ref{eq.red of SS-D}) is not algebraically equal to that of
 (\ref{eq.SABM-M}). Are these 
 two equations still consistent with each other ?  We will be able to answer
 to this question in the affirmative. 

Let us consider
\beqn
{\cal I}_{\rm R}={\cal D}_f^{\al}{\rm D}_{\al}\de(\eta-\te)\de(\bet-\bte)f(z)
\eeqn
where 
\beqn
f(z)&\equiv&\lan T^r W_S[C_{\zp z}]T^r\de^4(y-\yp)W_S[C_{z\zp}]\ran\,, \non\\
{\rm D}_{\al}&=&\frac{\pal}{\pal \eta^{\al}}-2i(\sig^a\bet)_{\al}
                   \frac{\pal}{\pal y^a}\,,\non\\
{\rm and}~~~~\bar{\rm D}_{\dal}&=&-\frac{\pal}{\pal \bet^{\dal}}\,,\non
\eeqn
and ${\cal D}_f^{\al}$ are given by (\ref{eq.Df}).
( Recall that we use the same notation for $z$ and $\zp$,
$z=(y,\eta,\bet),\,\,\zp=(\yp,\te,\bte)$  as before.)
After some calculation, we find that
${\cal I}_{\rm R}$ can be decomposed into
\beqn
{\cal I}_{\rm R}&=&{\cal I}_{\rm W-Z}+{\cal I}_{\rm d}\,, \label{eq.I_W+I_d}\\
{\cal I}_{\rm W-Z} 
&=&\{i(\eta-\te)\sig_a(\bet-\bte)dy^a
+\bte\bsi^b\sig_a\bet\de(\eta-\te)dy^a\pal_b\non\\
& &-2\bet^2\{d\eta(\eta-\te)+id\eta\sig^a\bte\de(\eta-\te)\pal_a\}f(z)\non\\
& &-\im{2}\bet^2\de(\eta-\te)dy^a\lan T^r W_S[C_{\zp z}]\de^4(y-\yp)
[\te\bsi_a{\cal W},T^r]W_S[C_{z\zp}]\ran\,,   \label{eq.I_W}  \\
{\cal I}_{\rm d}
&=&\bte^2\{2d\eta(\eta-\te)+\de(\eta-\te)dy^a\pal_a\}f(z)  \non\\
& &-\bte^2\bet^2\de(\eta-\te)\lan T^r W_S[C_{\zp z}]\de^4(y-\yp)
[d\eta{\cal W},T^r]W_S[C_{z\zp}]\ran\,. \label{eq.I_d}
\eeqn

In these calculations, ${\cal W}_{\al}$ comes from two origins. 
Let us consider the derivative $\bar{\rm D}_{\dal}\pal_a$ with this order 
and act on $f(z)$. First $\pal_a$ acts on $f(z)$, which generates
${\cal A}_a$  through $W_{S}$ in $f(z)$.
  The subsequent action of $\bar{\rm D}_{\dal}$ generates
 $\bar{\rm D}_{\dal}{\cal A}_a=-{\cal F}_{a\dal}$ and ${\cal W}_{\al}$.
( Note that $\pal_a\bar{\rm D}_{\dal}f(z)=0$ because
 $\bar{\rm D}_{\dal}W_S=0$.) The
${\cal W}_{\al}$ also appears when the derivative $\frac{\pal}{\pal\eta^{\al}}$
 acts on $f(z)$. This action yields
 ${\cal A}_{\underline{\al}}$ which has the Einstein index.
 From the definition (\ref{eq.Ay}), ${\cal A}_{\underline{\al}}$ can be
 expressed with ${\cal A}_a$ and ${\cal A}_{\al}$ by
\beqn
{\cal A}_{\underline{\al}}={\cal A}_{\al}-2i(\sig^a\bet)_{\al}{\cal A}_a\,.\non
\eeqn
Using the explicit expressions for ${\cal A}_{\al}$ and ${\cal A}_a$ in the 
Wess-Zumino gauge, we find
\beqn
{\cal A}_{\underline{\al}}
=\bet^2\{-i\lam_{\al}(y)+\eta_{\al}{\rm D}(y)
-i(\sig^{ab}\eta)_{\al}v_{ab}+\eta^2(\sig^a D_a\bla)_{\al} \}\,.  \non
\eeqn
Comparing this with (\ref{ex.W_(al)}), we find the following relations between 
 ${\cal A}_{\underline{\al}}$ and ${\cal W}_{\al}$:
\beqn
{\cal A}_{\underline{\al}}=\bet^2{\cal W}_{\al}\,.
\eeqn
The ${\cal W}_{\al}$ appears when $\frac{\pal}{\pal\eta^{\al}}$ acts on $f(z)$.

Returning to (\ref{eq.I_W+I_d}), we consider taking the trace of the 
first term ${\cal I}_{\rm W-Z}$ and integrate this along the closed loop $C$,
$I_{\rm W-Z}\equiv\oint\,tr{\cal I}_{\rm W-Z}$. 
Since ${\cal W}_{\al}$ is obtained by  the area derivative on the 
Wilson-loop $W_S[C_{\zp z}]$ or $W_S[C_{z\zp}]$, $I_{\rm W-Z}$ can be written 
 with  the super Wilson-loop and its area derivative alone  as 
\beqn
I_{\rm W-Z}
&=&\oint\{i(\eta-\te)\sig_a(\bet-\bte)dy^a
+\bte\bsi^b\sig_a\bet\de(\eta-\te)dy^a\pal_b \non\\
& &-2\bet^2d\eta(\eta-\te)-2i\bet^2d\eta\sig^a\bte\pal_a \}
\lan tr\,W_S[C_{\zp z}]\de^4(y-\yp) tr\,W_S[C_{z\zp}] \ran \non\\
& &+2\oint \1{8}
\bet^2\de(\eta-\te)\de^4(y-y^{\prime})dy^a(\bsi^b\sig_a)^{\dal}\non\\
& &\times \left\{\lan(\2{\Sig^{b\dal}(z)} tr\,W_S[C_{z^{\prime}z}])
                       tr\,W_S[C_{zz^{\prime}}]\ran
                    -\lan tr\,W_S[C_{z^{\prime}z}]
            (\2{\Sig^{b\dal}(z)} tr\,W_S[C_{zz^{\prime}}])\ran \right\}\non
\eeqn
This $I_{\rm W-Z}$ is 
exactly  the right hand side of the supersymmetric Schwinger-Dyson 
equation which we have obtained in the previous section.
  We find
\beqn
& &\1{8g^2}\ep_{\al\be}\bsi^{a\dal\be}{\rm D}^{\al}\2{\Sig^{a\dal}(\zp)}
\lan tr\,W_S[C]\ran  \non\\
&=&\oint tr\,{\cal I}_{\rm R}-\oint tr\,{\cal I}_{\rm d}  \non\\
&=&\oint {\cal D}^{\al}_f {\rm D}_{\al}\de(z-\zp)
\lan tr\,T^r W_S[C_{\zp z}]T^rW_S[C_{z\zp}] \ran-I_{\rm d}\,,\non\\
& &I_{\rm d}\equiv\oint  tr\,{\cal I}_{\rm d}\,.   \label{eq.SS-D2}
\eeqn
In the abelian reduction of the above equation, the first term of the 
right hand side is the same as that of eq.~(\ref{eq.SABM-M}). 

This fact indicates that $I_{\rm d}$ must identically be zero
 if the nonabelian equation is consistent with the abelian case.
  This is the case in fact.  From the relation 
${\cal A}_{\underline{\al}}=\bet^2{\cal W}_{\al}$ and (\ref{eq.sderivative2}),
\beqn
I_{\rm d}
&=&\1{2}\bte^2\oint \{dy^a\frac{\pal}{\pal y^a}
+d\eta^{\al}\frac{\pal}{\pal \eta^{\al}} \}\de(\eta-\te)\de^4(y-\yp)\non\\
& &\hspace*{1cm}\times\lan tr\,T^rW_S[C_{\zp z}]T^r W_S[C_{z\zp}]\ran \non\\
&=&\1{2}\bte^2\oint \{dy^a\frac{\pal}{\pal y^a}
+d\eta^{\al}\frac{\pal}{\pal \eta^{\al}}
+d\bet^{\dal}\frac{\pal}{\pal\bet^{\dal}} \}\de(\eta-\te)\de^4(y-\yp)\non\\
& &\hspace*{1cm}\times\lan tr\,T^rW_S[C_{\zp z}]T^r W_S[C_{z\zp}]\ran \non\\
&=&\1{2}\bte^2\oint dz^M\frac{\pal}{\pal z^M}
\de(\eta-\te)\de^4(y-\yp)\lan tr\,T^rW_S[C_{\zp z}]T^r W_S[C_{z\zp}]\ran\,,
                                                                       \non\\
& &
\eeqn
where we have used $\bar{\rm D}_{\dal}W_S=-\frac{\pal}{\pal\bet^{\dal}}W_S=0$ 
in the second equality. 
 This equation is the contour integral along the closed loop $C$ in superspace
  of the total derivative  and must vanish.

This identity is related to the restricted supergauge transformations which we 
have explained in Appendix C. Consider
\beqn
{\cal Z}[C_{\zp\zp}]
=\lan W_S[C_{\zp\zp}] \ran
=\int[d{\rm V}] {\rm e}^{i{\rm S}_{SYM}}W_S[C_{\zp\zp}]\,.
\eeqn  
Here we do not take the trace of $W_S[C_{\zp\zp}]$ and the integrand should be 
understood in the sense of matrix elements.
 Under the restricted supergauge transformations, the 
one-form ${\cal A}$ transforms as
\beqn
\de_{\Lambda}{\cal A}_M
=-\pal_M\Lambda+[{\cal A}_M,\Lambda]\,.
\eeqn
Using (\ref{eq.variationW_S}), we obtain
\beqn
0&=&\oint \lan W_S[C_{\zp z}]\de_{\Lambda}{\cal A}(z)W_S[C_{z\zp}]\ran \non\\
 &=&-\oint dz^M\frac{\pal}{\pal z^M}
    \lan W_S[C_{\zp z}]\Lambda(z)W_S[C_{z\zp}]\ran \,.
\eeqn
In terms of $(y,\eta,\bet)$,
$\frac{\pal}{\pal z^{M=\dal}}=\frac{\pal}{\pal\bet^{\dal}}=-\bar{\rm D}_{\dal}$
 and $\bar{\rm D}_{\dal}W_S=0$ as well as 
$\bar{\rm D}_{\dal}\Lambda=0$. We find that the above identity becomes 
\beqn 
0=\oint\{dy^a\frac{\pal}{\pal y^a}+d\eta^{\al}\frac{\pal}{\pal\eta^{\al}}\}
       \lan W_S[C_{\zp z}]\Lambda(z)W_S[C_{z\zp}]\ran\,.\label{eq.identity2}
\eeqn
 Acting   on eq.~(\ref{eq.identity2})
 the functional derivative $\2{\Lambda^{(r)}(\zp)}$, which satisfies 
\beqn
\2{\Lambda^{(r)}(\zp)}\Lambda(z)
=\1{4}\bar{\rm D}^{\prime2}\de(z-\zp)T^r
=\1{4}\bar{\rm D}^2\de(z-\zp)T^r \,, \non
\eeqn
 we find 
\beqn
0
&=&\1{4}\oint\{dy^a\frac{\pal}{\pal y^a}
+d\eta^{\al}\frac{\pal}{\pal\eta^{\al}}\}
\lan W_S[C_{\zp z}]\{\bar{\rm D}^{\prime2}\de(z-\zp)\}T^rW_S[C_{z\zp}]\ran
                                                                        \non\\
&=&\oint\{dy^a\frac{\pal}{\pal y^a}
+d\eta^{\al}\frac{\pal}{\pal\eta^{\al}}\}
\lan W_S[C_{\zp z}]\de(\eta-\te)\de^4(y-\yp)T^rW_S[C_{z\zp}]\ran\,.\non
\eeqn
Multiplying this equation by the generator $T^r$ and taking its trace, we see 
that the right hand side is equal to $I_{\rm d}$.

Inserting $I_{\rm d}=0$ into (\ref{eq.SS-D2}), multiplying the both sides
 by the gauge volume which we have thrown away in taking the Wess-Zumino gauge 
condition, we find the final form of the nonabelian supersymmetric 
Schwinger-Dyson equation for the super Wilson-loop:
\beqn
& &\1{8g^2}\ep_{\al\be}\bsi^{a\dal\be}{\rm D}^{\al}\2{\Sig^{a\dal}(\zp)}
\lan tr\,W_S[C]\ran  \non\\
&=&\oint{\cal D}^{\al}_f {\rm D}_{\al}\de(z-\zp)
\lan tr\,W_S[C_{\zp z}]tr\,W_S[C_{z\zp}] \ran \,,\non\\
& &{\cal D}_f^{\al}
=-e^a\frac{i}{4}\bar\sig_a^{\dal\al}\bar{\rm D}_{\eta\dot{\al}}+e^{\al}\,,
\non\\
& &\lan\,\cdots\,\ran =\int[d{\rm V}]\,{\rm e}^{i{\rm S}_{SYM}}\,\cdots\,,   
                                                           \label{eq.SS-Dfinal}
\eeqn
where we have used the closure property (\ref{eq.38}) of $U(N_c)$. 
This expression is manifestly invariant under the restricted supergauge 
transformations. 
The final equation obtained is the first among the infinite number of
  the Schwinger-Dyson equations which exhaust the dynamics of 
 supersymmetric Yang-Mills theory. At the same time, it exhibits
 string dynamics suggested by the theory.  In fact, the right hand side tells
 us that  the loop $C$ splits into two parts by the dynamics if two points
 on the loop coalesce at $z^{\prime}$; this latter condition comes from
 the delta function.
 
  Let us now derive  the rest of the Schwinger-Dyson equations.  This time,
   not only splitting of a loop but also  joining of  two loops  can take
 place   from the original configuration  
   $< \prod_{i=1}^{n} tr W_S[C^{(i)}] >$.  The functional derivative
 $\2{{\rm V}^{(r)}_{\rm mod}(z^{\prime})}{\cal A}(z)$
  $\; z^{\prime} \in C^{(j)}$,
 which appeared in the derivation of our equation in the previous section,
 now acts on every loop including the original one $C^{(j)}$.
 We find
\beqn
& &\1{8g^2}\ep_{\al\be}\bsi^{a\dal\be}{\rm D}^{\al}\2{\Sig^{a\dal}(\zp)}
\lan\prod_{i=1}^n\,trW_S[C^{(i)}]\ran\non\\
& &=\sum_{j=1}^n\te(\zp\in C^{(j)})\lan\,
\left(\prod_{i\neq j}^n\,tr W_S[C^{(i)}]\right)
\oint_{C^{(j)}}{\cal D}^{(j)\al}_f{\rm D}^{(j)}_{\al}\de(z_j-\zp)
trW_S[C^{(i)}_{\zp z_j}]trW_S[C^{(i)}_{z_j\zp}]\,\ran \non\\
& &+\sum_{j=1}^n\te(\zp\in C^{(j)})\sum_{k\neq j}^n
\lan\,\left(\prod_{i\neq j,k}^n\,trW_S[C^{(i)}]\right)
\oint_{C^{(k)}}{\cal D}^{(k)\al}_f{\rm D}^{(k)}_{\al}\de(z_k-\zp)
trW_S[C^{(k)}_{z_k z_k}+C^{(j)}_{\zp\zp}]\,\ran\,.\non\\
& &\label{eq.n-p SS-D}
\eeqn
 Here, the step function $\theta( z^{\prime} \in C^{(j)})$ indicates the case
 that $z^{\prime} $ is on $C^{(j)}$.
  As before, the second line of this equation
 describes the splitting of an individual loop
 into two if two points on the loop coalesce at $z^{\prime}$.
  The third line describes  the joining of two loops into one in the case where
 the point $z^{\prime}$ is shared by the two loops.  This latter condition
 is again a consequence of the delta function.

 As for the large $N_c$ limit with
 $g_c^2\equiv N_c g^2$ kept finite, we obtain 
 the supersymmetric extension 
 of the Migdal-Makeenko equation   which is summarized  in Appendix B:
\beqn
& &\1{8g_c^2}\ep_{\al\be}\bsi^{a\dal\be}{\rm D}^{\al}\2{\Sig^{a\dal}(\zp)}
     {\bf W}_S[C]  \non\\
&=&\oint{\cal D}^{\al}_f {\rm D}_{\al}\{\de(z-\zp)
 {\bf W}_S[C_{\zp z}]{\bf W}_S[C_{z\zp}]\}  \,,\non\\
& &{\bf W}_S[C_{z_1z_2}]\equiv \lan\1{N_c}tr\,W_S[C_{z_1z_2}]\ran 
=\int[d{\rm V}]\,{\rm e}^{i{\rm S}_{SYM}}tr\,W_S[C_{z_1z_2}]
                                   \,.      \label{eq.SM-Mfinal}
\eeqn
  The dynamics is then formally contained in the one-point average\footnote{
 There may be a subtlety in taking the large $N_{c}$ limit
 in the case of supersymmetric gauge theories.  This is related to
 the existence of degenerate vacua and the validity of the cluster property.}.

Eqs.~(\ref{eq.SS-Dfinal}) and (\ref{eq.SM-Mfinal}) are both manifestly 
supersymmetric. The supersymmetry transformations are understood as
 coordinate transformations in superspace:
\beqn
z=(y^a,\eta,\bet)\longrightarrow
\zp=(y^a+2i\eta\sig^a\bar{\xi},\eta+\xi,\bar{\eta}+\bar{\xi})\,,
                                                   \label{eq.coortr}
\eeqn
where $\xi$ and $\bar{\xi}$ denote the infinitesimal Grassmann parameters for
 the transformation. The derivatives ${\rm D}_A$ having flat indices are 
invariant under the coordinate transformations; they commute or 
anticommute with the differential operator ${\rm Q}$, $\bar{\rm Q}$ .
 The bases $e^A$ are 
invariant under the coordinate transformations as well,
\beqn
& &e^a=dy^a-2id\eta\sig^a\bet \non\\
&\longrightarrow&
e^{\prime a}=d\yp-2id\eta^{\prime}\sig^a\bet^{\prime} \non\\
& &\hspace*{1.5cm}=d(y^a+2i\eta\sig^a\bar{\xi})-2id\eta\sig^a(\bet+\bar{\xi})
                                                                \non\\
& &\hspace*{1.5cm}=e^a\,, \non\\
& &e^{\al}\longrightarrow e^{\al}\,,\,\,\,\,e^{\dal}\longrightarrow e^{\dal}\,.
\eeqn
The delta function $\de(z-\zp)=\de(\eta-\te)\de(\bet-\bte)\de^4(y-\yp)$ 
is obviously invariant.

  We also obtain
\beq
\2{\Sig^{\al\be}(\zp)}\lan tr\,W_S[C]\ran = 0\,,\;
\2{\Sig^{\al\dal}(\zp)}\lan tr\,W_S[C]\ran = 0\,,\;
\2{\Sig^{\dal\dbe}(\zp)}\lan tr\,W_S[C]\ran = 0\,.    \label{eq.flatSS-D}
\eeq
  from eq.~(\ref{eq.flatness}).
 These equations are identities  in the original variables but
should be treated as constraints   as soon as
  we employ the super Wilson-loop as a fundamental variable. 

In the large $N_c$ limit (\ref{eq.flatSS-D}) becomes 
\beq
\2{\Sig^{\al\be}(\zp)}{\bf W}_S[C]=0\,,\;
\2{\Sig^{\al\dal}(\zp)}{\bf W}_S[C] =0\,,\;
\2{\Sig^{\dal\dbe}(\zp)}{\bf W}_S[C] = 0\,.    \label{eq.flatSM-M}
\eeq

\section{Solution in the abelian case}
In this section, we solve the supersymmetric abelian Schwinger-Dyson equation. 
  Our main objective here is to check that  our nonlinear functional
 equation does contain the desirable nontrivial solutions 
  at least in the linearized approximation.
Start with
\beqn
& &\1{8g^2}\ep_{\al\be}\rD^{\al}\bsi^{a\dal\al}
\2{\Sig^{a\dal}(\zp)}\ln{\bf W}_S[C]
=\oint{\cal D}_f^{\al}\rD_{\al}\de(z-\zp)\,,\label{eq.solSM-M}\\
& &\hspace*{2cm}
\2{\Sig^{\al\be}(\zp)}\ln{\bf W}_S[C]=0 \,,\non\\
& &\hspace*{2cm}
\2{\Sig^{\dal\dbe}(\zp)}\ln{\bf W}_S[C]=0 \,,\non\\
& &\hspace*{2cm}
\2{\Sig^{\al\dal}(\zp)}\ln{\bf W}_S[C]=0 \,,\label{eq.solflat}
\eeqn
  which are respectively  (\ref{eq.SABM-M})  and (\ref{eq.flatSM-M}) 
 divided  for each side  by ${\bf W}_S[C]$.
These are the first order approximation to  the perturbative expansion of the 
supersymmetric nonabelian Schwinger-Dyson equation.

We now take an ansatz for the $\ln{\bf W}_S[C]$:
\beqn
\ln{\bf W}_S[C]=\oint e^A\oint e^{\prime B}\cP_{BA}(z,\zp)\,,\non\\
e^A=e^A(s)\,,\,\,\,z=z(s)\,,\;e^{\prime B}=e^B(t)\,,\,\,\,\zp=z(t)\,,\\
\left( \ln{\bf W}_S[C_{z_f z_i}]=\int_{z_i}^{z_f} e^A\int_{z_i}^{z_f}
 e^{\prime B}\cP_{BA}(z,\zp) \right)     \label{eq.anz}
\eeqn
where $\oint$ denotes the contour integration along the closed loop $C$. 
We use $s$ and $t$ to parameterize the loop $C$. 
 Exchanging $s$ and $t$, we find the relation
\beqn
\cP_{BA}(z,\zp)=(-)^{|A||B|}\cP_{AB}(\zp,z)\,.\label{eq.propanz}
\eeqn
Since the system given by (\ref{eq.solSM-M}) and (\ref{eq.solflat}) 
takes a manifestly supersymmetric form and
 the bases $e^A$  are invariant under the supertransformations, 
 $\cP_{BA}(z,\zp)$  must take the following form:
\beqn
\cP_{BA}(z,\zp)
=\cP_{BA}(y^a-y^{\prime a}-i(\eta-\eta^{\prime})\sig^a(\bet+\bet^{\prime}),
                   \eta-\eta^{\prime},\bet-\bet^{\prime}) =
\cP_{BA}(u^a,\tau,\bta)\,,\non
\eeqn
where we have introduced the new coordinates $u^a,\,\tau$ and $\,\bta$
  respectively by
\beqn
u^a \equiv y^a-y^{\prime a}-i(\eta-\eta^{\prime})\sig^a(\bet+\bet^{\prime})
\,,\;
\tau \equiv \eta-\eta^{\prime}\,,\;
\bta \equiv \bet-\bet^{\prime}\,.  \label{new coor}
\eeqn
These new coordinates are invariant under the supersymmetry
 transformations  (\ref{eq.coortr}).
  Useful formulas for
$\rD_A(y,\eta,\bte)$ or $\rDp_A(\yp,\eta^{\prime},\bet^{\prime})$  in terms of
$u^a,\,\tau,$ and $\,\bta$  are
\beqn
\rD_A
&=&\left\{
\begin{array}{l}               
\frac{\pal}{\pal y^a}=\frac{\pal}{\pal u^a}\,\,,  \\
\rD_{\al}=\frac{\pal}{\pal \eta^{\al}}
+2i(\sig^a\bet)_{\al}\frac{\pal}{\pal y^a}
=\frac{\pal}{\pal \tau^{\al}}+i(\sig^a\bta)_{\al}\frac{\pal}{\pal u^a}\,\,,\\
\brD_{\dal}=-\frac{\pal}{\pal \bet^{\al}}
=-\frac{\pal}{\pal \bta^{\dal}}-i(\tau\sig^a)_{\dal}\frac{\pal}{\pal u^a}\,\,.
\end{array}\right.   \non\\
\rDp_A
&=&\left\{  
\begin{array}{l}               
\frac{\pal}{\pal {\yp}^a}=-\frac{\pal}{\pal u^a}\,\,,  \\
\rDp_{\al}=\frac{\pal}{\pal \eta^{\prime\al}}
+2i(\sig^a\bet^{\prime})_{\al}\frac{\pal}{\pal {\yp}^a}
=-\frac{\pal}{\pal \tau^{\al}}+i(\sig^a\bta)_{\al}\frac{\pal}{\pal u^a}\,\,,\\
\brDp_{\dal}=-\frac{\pal}{\pal \bet^{\prime\al}}
=\frac{\pal}{\pal \bta^{\dal}}-i(\tau\sig^a)_{\dal}\frac{\pal}{\pal u^a}\,\,.
\end{array}\right.
\eeqn
Of course $\rD_A$ and $\rDp_A$ commute or anticommute with each other.

Under an infinitesimal deformation of the loop $C$, we obtain
\beqn
& &\de\ln{\bf W}_S[C]\non\\
& &=2\oint e^{\prime B}\de e^{\prime A}[\rDp_A\oint e^C\cP_{CB}(\zp,z)
-(-)^{|A||B|}\rDp_B\oint e^C\cP_{CA}(\zp,z)]\non\\
& &+4i\oint(e^{\prime\al}\de e^{\prime\dal}+e^{\prime\dal}\de e^{\prime\al})
\sig^a_{\al\dal}\oint e^A\cP_{Aa}(\zp,z) \;\;,     \label{eq.infinidef}
\eeqn
where we used (\ref{eq.propanz}) and $\de e^{\prime A}$ is defined by
\beqn
\de e^{\prime A}=\de z^M(t)e_M\,^A(t)\,.\non
\eeqn
The last term of the right hand side of (\ref{eq.infinidef}) appears because 
we express it with the bases $e^A$.
The additional anzatz we take for the solution reads
\beqn
\cP_{A\dal}=0\non
\eeqn
 in  $\ln {\bf W}_S[C]$.
This ansatz is consistent with the condition (\ref{eq.chirality})
\beqn
0=\brD_{\dal}\ln{\bf W}_S[C_{z_f z_i}]
\,\,=2\int_{z_i}^{z_f}e^{\prime B}\cP_{B\dal}(z_f\,\,{\rm or}\,\,z_i,\,\zp)\;.
\eeqn

Let us return to (\ref{eq.infinidef}). The contour integral 
$\oint e^{\prime B}\de e^{\prime A}$ 
in (\ref{eq.infinidef}) represents an infinitesimal area element which has
flat 
indices. Therefore we get, from (\ref{eq.infinidef}), the following
 constraints given by (\ref{eq.solflat})
\beqn
0&=&\2{\Sig^{\al\be}(\zp)}\ln{\bf W}_S[C]\non\\
&=&2\oint e^a\{\rDp_{\al}\cP_{a\be}(\zp,z)+\rDp_{\be}\cP_{a\al}(\zp,z)\}\non\\
& &-2\oint e^{\gamma}\{\rDp_{\al}\cP_{\gamma\be}(\zp,z)
+\rDp_{\be}\cP_{\gamma\al}(\zp,z)\}\,,\label{eq.const1}\\
0&=&\2{\Sig^{\dal\al}(\zp)}\ln{\bf W}_S[C]\non\\
&=&2\oint e^a\{\brDp_{\dal}\cP_{a\al}(\zp,z)
+2i\sig^b_{\al\dal}\cP_{ab}(\zp,z)\}\non\\
& &-2\oint e^{\gamma}\{\brDp_{\dal}\cP_{\gamma\al}(\zp,z) 
+2i\sig^b_{\al\dal}\cP_{\gamma b}(\zp,z)\}\,,\label{eq.const2}
\eeqn
where we used (\ref{eq.propanz}) and the condition $\cP_{A\dal}=0$. The 
equation originating from the second one of (\ref{eq.solflat}) is trivially 
satisfied  as $\cP_{A\dal}=0$.

 From (\ref{eq.const1}), we obtain
\beqn
0&=&\rDp_{\al}\cP_{a\be}(\zp,z)+\rDp_{\be}\cP_{a\al}(\zp,z)\,,\non\\
0&=&\rDp_{\al}\cP_{\gamma\be}(\zp,z)+\rDp_{\be}\cP_{\gamma\al}(\zp,z)\,,
                                                          \label{eq.const1P}
\eeqn
and $\cP_{a\al},\,\cP_{\gamma\al}$ are written  as 
\beqn
\cP_{a\al}(\zp,z)=\rDp_{\al}\cP_a(\zp,z)\,,\;
\cP_{\gamma\al}(\zp,z)=\rDp_{\al}\cP_{\gamma}(\zp,z)\,.\label{eq.pure}
\eeqn
for $\cP_a(\zp,z),\,\cP_{\gamma}(\zp,z)$.
 From (\ref{eq.const2}) and (\ref{eq.pure}), we obtain 
\beqn
\cP_{ab}(\zp,z)=-\im{4}\brDp\bsi_b\rDp\cP_a(\zp,z)\,,\;
\cP_{\gamma b}(\zp,z)= -\im{4}\brDp\bsi_b\rDp\cP_{\gamma}(\zp,z)\,.\non
\eeqn
Considering the above forms of $\cP_{ab},\,\cP_{\gamma b}$ and also 
(\ref{eq.propanz}), we can express all $\cP$'s with  single function 
$\cP(\zp,z)$:
\beqn
\cP_b(\zp,z)&=&-\im{4}\brD\bsi_b\rD\cP(\zp,z)\,,\non\\
\cP_{\gamma}(\zp,z)&=&\rD_{\gamma}\cP(\zp,z)\,,\non\\
\cP_{ab}(\zp,z)&=&(\im{4})^2\brD\bsi_a\rD\brDp\bsi_b\rDp\cP(\zp,z)\,,\non\\
\cP_{\gamma b}(\zp,z)&=&-\im{4}\brDp\bsi_b\rDp\rD_{\gamma}\cP(\zp,z)\,,\non\\
\cP_{\gamma\al}(\zp,z)&=&\rDp_{\al}\rD_{\gamma}\cP(\zp,z)\,.\label{eq.pure2}
\eeqn
Here $\cP(\zp,z)$ must satisfy the condition 
\beqn
\cP(\zp,z)=\cP(z,\zp)\,.      \label{eq.propP}
\eeqn
This function should  be
 a function of $u,\,\tau,\,\bta$ only.
 We see that $\cP$ must take the following form 
\beqn
\cP(\zp,z)=Q(u)+\tau^2\bta^2 R(u)\,. \label{eq.formP}
\eeqn

The remaining task is to determine the function $\cP(z,\zp)$ by 
considering equation (\ref{eq.solSM-M}). Using (\ref{eq.infinidef}) we obtain
\beqn
& & 2\oint e^a\bsi^{b\dal\al}\rDp_{\al}\brDp_{\dal}\cP_{ab}(\zp,z)\non\\
& &+2\oint e^{\be}\bsi^{a\dal\al}\rDp_{\al}\brDp_{\dal}\cP_{\be a}(\zp,z)
\non\\
&=&8g^2\oint\{{\cal D}_f^a\rD_{\al}\de(\tau)\de(\bta)\de^4(u)
+e^A\rD_A f(u,\tau,\bta)\}\,.\label{eq.of.P}
\eeqn
Here we add the last term $8g^2\oint e^A\rD_A f(u,\tau,\bta)$ to the right 
hand side of the Schwinger-Dyson equation. This term is auxiliary  as 
it has the form of total derivatives for the contour integral. This 
additional term is not necessary and we add it only for convenience. The 
source term ${\cal D}_f^{\al}\rD_{\al}\de(\tau)\de(\bta)\de^4(u)$ does not 
involve the basis $e^{\dal}$. Following this fact, we impose the condition
 that $f$ be  chiral. The $f(u,\tau,\bta)$ must then have the 
form
\beqn
f(u,\tau,\bta)
 ={\rm A}(u)+i\tau\sig^a\bta\pal_a{\rm A}(u)+\1{4}\tau^2\bta^2\Box{\rm A}(u)
+\tau^2{\rm F}(u) \;.
\eeqn

We now insert $\cP_{ab}(\zp,z),$ and $\cP_{\gamma b}(\zp,z)$
(see (\ref{eq.pure})) into (\ref{eq.of.P}).
Using the concrete form of 
$\cP$ given by (\ref{eq.formP}) and matching appropriate power of 
$\tau,\,\bta$, we get the following independent differential equations of the 
fields $Q(u),\,R(u)$ and the component fields of $f(u,\tau,\bta)$,
\beqn
& &\Box Q(u)-4R(u)=-4ig^2{\rm A}(u)\,,\non\\
& &\Box(\Box Q(u)-4R(u))=4ig^2\de^4(u)\,,\non\\
& &{\rm F}(u)=0\,.
\eeqn
Solving these equations, we find that 
\beqn
\cP(z,\zp)&=&2ig^2\Box^{-1}(\Box^{-1}\de^4(u)-C(u)) \non\\
          & &-\im{2}g^2\tau^2\bta^2(\Box^{-1}\de^4(u)+C(u))\,.
\eeqn
Here the  $C(u)$ is an arbitrary function and can be regarded as 
 gauge parameter. We  need not be concerned about the
 dependence of $\cP(z,\zp)$ on $C(u)$.
 Only $\cP_{AB}(\zp,z)$ are directly related to physical quantities
  and these are  related to $\cP(z,\zp)$ 
through the gauge independent combination $I(u)=\Box Q(u)-4R(u)$. Using the 
expressions (\ref{eq.pure2}), we obtain that
\beqn
\cP_{ab}(\zp,z)
&=&(\im{4})^2\{2g_{ab}-2i\tau\sig_a\bta\pal_b+2i\tau\sig_b\bta\pal_a\non\\
& &+2\ep_{abcd}\pal^c\tau\sig^d\bta
-\tau^2\bta^2(g_{ab}\Box-\pal_a\pal_b)\}I(u) \,,\non\\
\cP_{\al b}(\zp,z)
&=&-\im{4}\{(\sig_a\bta)_{\al}+\im{2}\bta^2(\sig^b\bsi_a\tau)_{\al}\pal_b\non\\
& &+i\tau_{\al}\bta^2\pal_a\}I(u)  \,,  \non\\
\cP_{\al \be}(\zp,z)&=&\1{2}\bta^2\ep_{\al\be}I(u)\,,  \non\\
I(u)&=&4ig^2\Box^{-1}\de^4(u)\,.
\eeqn
These results are equal to those obtained by 
carrying out the path integral for the super Wilson-loop average. 
So we can conclude that our Schwinger-Dyson equation is  nontrivial
 and provides
 no less information than the path integral method does. We should 
note that the equations given by (\ref{eq.solflat}) are as important as 
(\ref{eq.solSM-M}) in order to solve our Schwinger-Dyson equation.

\section{One-dimensional fermion along the loop and
 renormalization of the one-point super Wilson-loop average}

In the previous section, we obtained the abelian as well as the
nonabelian supersymmetric Schwinger-Dyson equations. These are, however,
written in terms of bare quantities  and must be renormalized in order to
obtain physical results.
For the ordinary non-supersymmetric Wilson-loop, the renormalization  of
 the one-point function has been already
 discussed in refs.~\cite{GerNe}-\cite{Ao}. The renormalization of the 
Wilson-loop is most transparent in terms of  the first quantized lagrangian
 minimally
  coupled to the gauge fields 
(see~\cite{IH,GerNe}):
\beqn
\cL_{path}=\int_0^1ds\,[i\bw(s)\dw(s)+g\bw(s)\dot{x}^a(s)v_a(x(s))w(s)]\,,
\label{effact}
\eeqn
where $w(s)$ is a one-dimensional fermion on a given path $x^a(s)$ and belongs
 to the fundamental representation of $SU(N_c)$ or $U(N_c)$.
 In this language,
 the path-ordered exponential is  represented as the two point function
\beqn
W[C]=[P\exp\{ig\int_0^1ds\dot{x}v_a\}]^{ij}
=\lan0|P\,w^i(0)\bw^j(1)|0\ran\,.  \label{Wilcor}
\eeqn
 The proof of this  equality becomes  obvious if one recognizes  that the
 right hand side of the above equation is a Green's function $G(s=1)$
 satisfying 
\beqn
 \left( i \frac{d}{ds} + g \dot{x}^{a}(s)v_{a}(s) \right) G(s) = i \delta(s)
 \delta_{i}^{j} \;\;.
\eeqn
 The solution of this equation under $G(0)=1$ is in fact the left hand side of
 (\ref{Wilcor}) with multiplication of the step function $\theta(s)$ implied.
Therefore, the renormalization problem of the Wilson-loop
 - a composite operator - becomes equivalent to that of
  the $w$-field in the lagrangian $\cL_{path}+\cL_{YM}$

We may expect that the renormalization problem of our super Wilson-loop can
 be handled in a similar way to  the non-supersymmetric case.
Let us define
\beqn
\cL_{path}=\int_0^1ds\,[i\bw(s)\dw(s)-i\bw(s)\cA(z(s))w(s)]\,,\label{Seff}
\eeqn
where $\cA(z)=e^A\cA_A$ is the one-form in superspace.
Using $\cL_{path}$, we can write the super path-ordered exponential in the
 form similar to (\ref{Wilcor}),
\beqn
W_S[C]=[P\exp\oint\cA(z(s))]^{ij}=\lan0|P\,w^i(0)\bw^j(1)|0\ran\,.
\eeqn
Let us demonstrate  the renormalization of the self-energy part of the 
$w$-field to one-loop order by using the dimensional regularization.
We will show that the pole 
term appearing in the self-energy part is cancelled by the local counterterm 
\beqn
\int_0^1ds\,i(Z_2-1)\bw(s)\dw(s)\,.
\eeqn
 Here $Z_2$ denotes the renormalization constant of the wave function  for
 the $w$ field.
 
At one-loop, all we have to do is to consider the leading 
of the vertex $\int\bw\cA w$ expanded in $\rV$. The ghost loop does 
not contribute to the one-loop self-energy part. Knowing these we can
 proceed without taking the Wess-Zumino gauge. 
The propagator for the $w$-field and  that for the superfield $\rV(z)$ are
 respectively
\beqn
& &\lan w^{(i)}(s_1)\bw^{(j)}(s_2)\ran=\de^{ij}\te(s_2-s_1)\non\\
& &\lan \rV^{(r)}(z)V^{(s)}(\zp)\ran\non\\
& &=-ig^2\de^{rs}\{-4(1-\al)\Box_{\rD}^{-2}(u)
+(1+\al)\tau^2\bta^2\Box_{\rD}^{-1}(u)\}\non\\
& &\equiv-ig^2\de^{rs}\rV(z,\zp)\,,   \label{propagator}
\eeqn
where $\Box_{\rD}^{-1}(u)$ and 
$\Box_{\rD}^{-2}(u)$ are defined respectively by the 
following forms
\beqn
\Box_{\rD}^{-1}(u)
&=&\int \frac{d^{\rD}k}{(2\pi)^{\rD}}\frac{{\rm e}^{-iku}}{-k^2}
=-\im{4\pi^{\rD/2}}\frac{\Gamma(\frac{\rD}{2}-1)}{u^{\rD-2}}\,,\non\\
\Box_{\rD}^{-2}(u)
&=&\int \frac{d^{\rD}k}{(2\pi)^{\rD}}\frac{{\rm e}^{-iku}}{(-k^2)^2}
=\im{16\pi^{\rD/2}}\frac{\Gamma(\frac{\rD}{2}-2)}{u^{\rD-4}}\,.
\eeqn
The variables $u,\tau$ and $\bta$ are the same ones as defined by 
(\ref{new coor}). Here $u$ is defined by 
\beqn
u=\sqrt{u^au_a}=\sqrt{-u_0^2+u_1^2+u_2^2+u_3^2}\,.
\eeqn
Following this definition of $u$, the Lorentz index $a$ runs only from 0 to 3.
The matrices $\sig^a$ or $\bsi^a$ are defined in 4-dimensional spacetime.

$\Box_{\rD}^{-1}(u)$ and $\Box_{\rD}^{-2}(u)$ satisfy 
\beqn
\Box\Box_{\rD}^{-2}(u)&=&(1+\ep)\Box_{\rD}^{-1}(u)\,,\non\\
\Box\Box_{\rD}^{-1}(u)&=&\de_{\rD}(u)\,,\non\\
\de_{\rD}(u)&=&i\frac{\Gamma(\rD/2)}{\pi^{\rD/2}}\ep u^{2\ep-4}\,,\non\\
\ep&=&\frac{4-\rD}{2}\,,\non
\eeqn
where $\Box$ denotes the four dimensional Klein-Gordon operator.

The propagator for $\rV$ is invariant under the four dimensional supersymmetry 
transformations given by (\ref{eq.coortr}) and is expressible in terms of
 $u,\tau$ and $\bta$.
Using this propagator and the vertex interaction $\int ds\,\bw\cA w$ 
in $\cL_{Seff}$,  we find
\beqn
i\Sig^{(1)}
&=&(i)^2\int_0^1ds_1\int_0^1ds_2\bw(s_1) T^rT^s w(s_2)\te(s_2-s_1)
\lan \cA^{(r)}(z)\cA^{(s)}(\zp)\ran  \non\\
&=&-ig^2(i)^2C_2(N_c)\int_0^1ds_1\int_0^1ds_2\bw(s_1)w(s_2)\te(s_2-s_1)\non\\
& &\times\{(\im{4})^2e^ae^{\prime b}\brD\bsi_a\rD\,\brDp\bsi_b\rDp\non\\
& &-\im{4}e^ae^{\prime \al}\brD\bsi_a\rD\,\rDp_{\al}
   -\im{4}e^{\al}e^{\prime a}\rD_{\al}\,\brDp\bsi_a\rDp\non\\
& &+e^{\al}e^{\prime \be}\rD_{\al}\rDp_{\be}\}\rV(z,\zp)\,,\non\\
C_2(N_c)&=&\frac{N_c^2-1}{N_c}\,\,\mbox{for gauge group}\,\,SU(N_c)\,,\,\,
N_c,\,\mbox{for gauge group}\,\,U(N_c)\,,
\eeqn
where we have taken into account that only the leading order in $\rV$ in the
 vertex contributes to the one-loop self energy as we mentioned above.

Let us evaluate the first term in $i\Sig^{(1)}$, which we denote by $iI_1$.
 The term $iI_1$ is given by
\beqn
iI_1
&\equiv&-i(\im{4})^2(ig)^2 C_2(N_c)\int_0^1ds_1\int_{s_1}^1ds_2\,\bw(s_1)w(s_2)
e^ae^{\prime b}\non\\
& &\times\{-16g_{ab}\Box_{\rD}^{-1}(u)+16i\tau\sig_b\bta\pal_a\Box_{\rD}^{-1}(u)
-16i\tau\sig_a\bta\pal_b\Box_{\rD}^{-1}(u)\non\\
& &-16\ep_{abcd}\tau\sig^c\bta\pal^d\Box_{\rD}^{-1}(u)
+4\tau^2\bta^2g_{ab}\de_{\rD}(u)\}k(\ep,\al)\,,\non\\ 
& &k(\ep,\al)=1+\1{2}\ep(1-\al)\,,\,\,\ep=\frac{4-\rD}{2}\,.\label{eq.I_1}
\eeqn
  We have used  (\ref{propagator}).
 The function $\de_{\rD}(u)$ is a well-defined distribution and does not have 
poles at $\rD=4$. The last term in the middle bracket does not contribute to
 the divergence.
The divergence comes from the point $s_1=s_2$. In order to estimate 
the behavior of the integrand of (\ref{eq.I_1}) around this point, 
  we expand as Taylor series
\beqn
& &e^{\prime a}=e^a(s_2)=e^a(s_1)+(s_2-s_1)de^a(s_1)+\cdots\,,\non\\
& &w(s_2)=w(s_1)+(s_2-s_1)\dw(s_1)+\cdots\,,\non\\
& &u^a=(s_1-s_2)e^a(s_1)-\1{2}(s_1-s_2)^2de^a(s_1)+\cdots \,,\non\\  
& &\tau=(s_1-s_2)\dot{\eta}(s_1)+\cdots\,,\non\\
& &\bta=(s_1-s_2)\dot{\bet}(s_1)+\cdots\,.\label{eq.expan}
\eeqn
The first term in $I_1$ reads
\beqn
& &-i(\im{4})^2(ig)^2C_2(N_c)\int_0^1 ds_1\int_{s_1}^1 ds_2
\bw(s_1)w(s_2)e^ae^{\prime b}
(-16)g_{ab}\Box_{\rD}^{-1}(u)\non\\
&=&16(\im{4})^2(ig)^2 C_2(N_c)
\1{4\pi^{\rD/2}}\Gamma(\frac{\rD}{2}-1)\int_0^1ds_1\int_{s_1}^1ds_2\,
|e|^{4-\rD}\{(s_1-s_2)^{2-\rD}\bw(s_1)w(s_1)\non\\
& &+\1{2}\frac{ede}{|e|^2}(\rD-4)(s_1-s_2)^{3-\rD}\bw(s_1)w(s_1)
+(s_1-s_2)^{3-\rD}\bw(s_1)\dw(s_1) \}k(\ep,\al) \non\\
& &+O((\rD-4))\,,\hspace*{2cm}|e|^2=e^ae_a\,.    \label{eq.I_1_t}
\eeqn
The term of $(s_1-s_2)^{2-\rD}$ does not yield any pole at $\rD=4$ 
after integrating over $s_2$.
The second term yields  a pole, but it gets multiplied by the factor $(\rD-4)$.
 The contribution to the pole at $\rD=4$ comes only from the last term of 
(\ref{eq.I_1_t}).
We find
\beqn
& &\1{8\pi^2}(ig)^2C_2(N_c)\int_0^1ds_1\,\bw(s_1)\dw(s_1)
\{-\frac{2}{4-\rD}-\gamma-\1{2}(1-\al)-\ln(\pi|e|^2(1-s_1)^2)\}\non\\
& &+({\rm finite\,\,term\,\,at\,\,D=4})\,.\label{eq.I_1_1}
\eeqn
Next we evaluate the second, third and forth terms in the middle bracket of 
the integrand of (\ref{eq.I_1}). By using (\ref{eq.expan}), we see that each of
these terms has poles but does not contribute to the self-energy part
 because of 
antisymmetry with respect to the indices $a$ and $b$.

We will estimate the second and the third terms in $i\Sig^{(1)}$,
 which we denote respectively by $iI_{2}$ and $iI_{3}$. Using the 
expansion (\ref{eq.expan}), the second term $iI_2$ is evaluated as 
\beqn
iI_2
&\equiv&i\im{4}(ig)^2C_2(N_c)\int_0^1ds_1\int_0^1ds_2\,\bw(s_1)w(s_2)
\te(s_2-s_1)
e^ae^{\prime \al}\brD\bsi_a\rD\,\rDp_{\al}V(z,\zp)\non\\
&=&i\im{4}(ig)^2C_2(N_c)\int_0^1ds_1\int_{s_1}^1ds_2\,\bw(s_1)w(s_2)
e^ae^{\prime \al}\{8(\sig_a\bta)_{\al}\Box_{\rD}^{-1}\non\\
& &-8i\tau_{\al}\bta^2\pal_a\Box_{\rD}^{-1}
-4i\bta^2(\sig^b\bsi_a\tau)_{\al}\pal_b\Box_{\rD}^{-1}\}k(\ep,\al)\non\\
&=&-\im{2\pi^{\rD/2}}(ig)^2C_2(N_c)\int_0^1ds_1\,\bw(s_1)w(s_1)|e|^{2-\rD}
\Gamma(\frac{\rD}{2}-1)\1{4-\rD}e^ae^{\al}\sig_{a\al\dal}e^{\dal}
(s_1-1)^{4-\rD}
\non\\
& &+({\rm finite\,\,term\,\,at\,\,\rD=4})\,.
\eeqn
Similarly, the third term $iI_3$ is evaluated as
\beqn
iI_3
&\equiv&i\im{4}(ig)^2C_2(N_c)\int_0^1ds_1\int_0^1ds_2\,\bw(s_1)w(s_2)\te(s_2-s_1)
e^{\prime a}e^{\al}\brDp\bsi_a\rDp\,\rD_{\al}V(z,\zp)\non\\
&=&\im{2\pi^{\rD/2}}(ig)^2C_2(N_c)\int_0^1ds_1\,\bw(s_1)w(s_1)|e|^{2-\rD}
\Gamma(\frac{\rD}{2}-1)\1{4-\rD}e^ae^{\al}\sig_{a\al\dal}e^{\dal}
(s_1-1)^{4-\rD}
\non\\
& &+({\rm finite\,\,term\,\,at\,\,\rD=4})\,.
\eeqn
The  first line of $iI_3$ has an opposite sign to that of $iI_2$. We conclude 
 that  $iI_2+iI_3$ does not have a pole at $\rD=4$.
The integrand of the last term in $i\Sig^{(1)}$ includes only higher powers of 
$(s_1-s_2)$ than $4-\rD$ when expanded around $s_1$. It
does not yield a pole.
 
Putting all these together, we obtain 
\beqn
i\Sig^{(1)}
&=&\1{8\pi^2}(ig)^2C_2(N_c)\int_0^1ds_1\,\bw(s_1)\dw(s_1)
\{-\frac{2}{4-\rD}+\gamma-\1{2}(1-\al)-\ln(\pi|e|^2(1-s_1)^2)\}\non\\
& &+({\rm finite\,\,term\,\,at\,\,\rD=4})\,.
\eeqn
 At one-loop level, the self-energy part of the $w$-field
 is renormalized by the local counterterm 
\beqn
\int_0^1ds\,i(Z_2-1)\bw(s)\dw(s)\,,\non
\eeqn
  with $Z_2$ given by
\beqn
Z_2=1-\frac{g^2}{8\pi^2}C_2(N_c)\{-\frac{2}{4-\rD}+\gamma-\1{2}(1-\al)\}\,.
                                                                  \label{Z_2}
\eeqn
This one-loop renormalization constant $Z_2$ is the same one as in the 
non-supersymmetric case.

\section{Discussion}

  We have already summarized the results from our investigation
 in introduction. Let us
   briefly  discuss  a few points which we would like to pursue.
  In section seven,  we considered  the renormalization of one-point
function  for our super Wilson-loop and carried out the explicit
 one-loop computation.  We should, however, consider the renormalization
 of our Schwinger-Dyson equation (\ref{eq.SM-Mfinal}) as well.
   One may expect that this could be done by
 smearing the delta function $\de(z-\zp)$ appearing in (\ref{eq.SM-Mfinal})
 by a heat kernel.  Another obvious direction  is extension of our work to
  the $N=2$ case.  We have made some preliminary investigations and
 hope to be able to report on this. A wealth of results will be waiting
  on this avenue in connection with \cite{SW,Int}.

\appendix
\section{A. ~~Wilson-loop and area derivative}
  
  We represent the gauge field  by 
$v_a(x)=\sum_r v^{(r)}_a(x)
T^{(r)}$ where $T^{(r)}$ are the generators of the gauge group.
 Using the spacetime curve $C_{xy}$, we define the Wilson-loop as follows,
\begin{eqnarray}
W[C_{xy}]=P\exp\{-\frac{i}{2} \int_y^x dw^a v_a(w) \}\,\,,  \label{eq.PWilson}
\end{eqnarray}
where the capital $P$ denotes the usual path ordered product. 
The field strength in this notation reads 
\[v_{ab}=\partial_a v_b - \partial_b v_a + \frac{i}{2} [v_a , v_b] \,\,. \]
  We simply list
 \begin{eqnarray}
 \frac{\partial}{\partial y^a}W[C_{xy}] &=& +\frac{i}{2} W[C_{xy}]v_a(y)
                                                      \,,\nonumber\\
 \frac{\partial}{\partial x^a}W[C_{xy}] &=& -\frac{i}{2} v_a(x)W[C_{xy}]\,.
                                                      \label{eq.derivativeW} 
 \end{eqnarray}

 Let us now consider two parameters $s$,
$t$, and a function $u^a(s,t)$ which maps these parameters into spacetime.
 For convenience we take the condition $u^a(0,0)=x^a$. We consider two small
 curves $\triangle C_1$ and $\triangle C_2$ defined by $u^a(s,t)$ as follows:
 \begin{eqnarray}
 \triangle C_1 &:& u(0,0)\stackrel{{\rm fixing}\, t = 0}{\longrightarrow}
  u(\delta s,0)\stackrel{{\rm fixing}\, s = \delta s}{\longrightarrow}
  u(\delta s,\delta t)\,, \\ 
 \triangle C_2 &:& u(0,0)\stackrel{{\rm fixing}\, s = 0}{\longrightarrow}
  u(0,\delta t)\stackrel{{\rm fixing}\, t = \delta t}{\longrightarrow}
  u(\delta s,\delta t) \,,
 \end{eqnarray}
where $\delta s$ and $\delta t$ are infinitesimal. Now we will evaluate the 
difference between $W[\triangle C_1+C_{xy}]$ and $W[\triangle C_2+C_{xy}]$,
 \begin{eqnarray}
 \delta W[C_{xy}]&\equiv&W[\triangle C_1+C_{xy} ] - W[\triangle C_2+C_{xy} ] 
                                                                  \nonumber\\
                    &=& \{W[\triangle C_1] - W[\triangle C_2] \}W[C_{xy}] \,,
 \end{eqnarray}
where $\delta W[C_{xy}]$ denotes that
 difference. 
Keeping the term up to the second order in $\delta s$ and $\delta t$, we get 
 \begin{eqnarray}
 W[\triangle C_1] - W[\triangle C_2] &=& -\frac{i}{2}P\{\int_{\triangle C_1}
  - \int_{\triangle C_2} \}dw^a v_a(w) \nonumber\\
   & & + \frac{1}{2}\big(-\frac{i}{2}\big)^2 
    P \{(\int_{\triangle C_1}dw^a v_a(w))^2 
     - (\int_{\triangle C_2}dw^a v_a(w))^2 \} \nonumber\\
   & & + O(\delta^3)\,.  \label{eq.C1-C2} 
 \end{eqnarray}
Calculating $\int_{\triangle C_1}dw^a v_a(w)$ of the right hand side of 
(\ref{eq.C1-C2}), 
we find that
 \begin{eqnarray}
 \int_{\triangle C_1}dw^a v_a(w)
  &=& \{\delta s \frac{\partial u^a}{\partial s} 
        + \delta t \frac{\partial u^a}{\partial t} \}v_a(x) \nonumber\\
  &+& \frac{1}{2}(\delta s)^2 \{\frac{\partial^2 u^a}{\partial s^2}v_a(x)
              + \frac{\partial u^a}{\partial s}\frac{\partial u^b}{\partial s}
                  \partial_a v_b(x) \} \nonumber\\
  &+& \frac{1}{2}(\delta t)^2 \{\frac{\partial^2 u^a}{\partial t^2}v_a(x)
               + \frac{\partial u^a}{\partial t}\frac{\partial u^b}{\partial t}
                \partial_a v_b(x) \} \nonumber\\
 &+& \delta s \delta t\quad\{\frac{\partial^2 u^a}{\partial s \partial t}v_a(x)
               + \frac{\partial u^a}{\partial s}\frac{\partial u^b}{\partial t}
               \partial_a v_b(x) \} \,\,,
 \end{eqnarray}
and $\int_{\triangle C_2}dw^a v_a(w)$ is given similarly as 
$\int_{\triangle C_1}dw^a v_a(w)$ with $\delta s$ and $\delta t$ exchanged. 
So we obtain  
 \begin{eqnarray}
  -\frac{i}{2}P\{\int_{\triangle C_1}
  - \int_{\triangle C_2} \}dw^a v_a(w)
 &=& -\frac{i}{2}\delta s \delta t\frac{1}{2}\{\frac{\partial u^a}{\partial s}
     \frac{\partial u^b}{\partial t}
     - \frac{\partial u^b}{\partial s}\frac{\partial u^a}{\partial t} \}
       (\partial_a v_b(x) - \partial_b v_a(x)) \,.\nonumber\\
 & &           \label{eq.first}
 \end{eqnarray}
Calculating $P (\int_{\triangle C_1}dw^a v_a(w))^2$ of the right hand side of 
(\ref{eq.C1-C2}), we get
 \begin{eqnarray}
 P (\int_{\triangle C_1}dw^a v_a(w))^2 
  &=& \delta s^2 \{ \frac{\partial u^a}{\partial s}v_a(x) \}^2
     + \delta t^2 \{ \frac{\partial u^a}{\partial t}v_a(x) \}^2 \nonumber\\
  &+& 2\delta s \delta t\frac{\partial u^a}{\partial s}
        \frac{\partial u^b}{\partial t}v_b(x)v_a(x)\,\,,
 \end{eqnarray}
and $P (\int_{\triangle C_2}dw^a v_a(w))^2$ is the same as 
$P (\int_{\triangle C_1}dw^a v_a(w))^2$ with $\delta s$ and $\delta t$ 
exchanged. 
So we obtain
 \begin{eqnarray}
  & &\frac{1}{2}\big(-\frac{i}{2}\big)^2 
   P \{(\int_{\triangle C_1}dw^a v_a(w))^2 
  - (\int_{\triangle C_2}dw^a v_a(w))^2 \} \nonumber\\
  & & = -\frac{1}{2}\big(-\frac{i}{2}\big)^2\delta s \delta t 
       \{\frac{\partial u^a}{\partial s}
       \frac{\partial u^b}{\partial t} - \frac{\partial u^b}{\partial s}
       \frac{\partial u^a}{\partial t} \}[v_a(x),v_b(x)] \,.\label{eq.second}
 \end{eqnarray}
From the equation (\ref{eq.first}) and (\ref{eq.second}), we see that 
 \begin{eqnarray}
  \delta W[C_{xy}] &=& \big(-\frac{i}{2}\big) \frac{1}{2} \delta s \delta t 
     \{\frac{\partial u^a}{\partial s}
       \frac{\partial u^b}{\partial t} - \frac{\partial u^b}{\partial s}
        \frac{\partial u^a}{\partial t} \}v_{ab}(x)W[C_{xy}] \,.
 \end{eqnarray}
Let us consider a closed curve $\triangle C \equiv \triangle C_1 
- \triangle C_2$.
We project $\triangle C$ to the $(x^a,x^b)$-plane and express this curve 
$\triangle C^{ab}$.
$\frac{1}{2} \delta s \delta t 
    \{\frac{\partial u^a}{\partial s}
      \frac{\partial u^b}{\partial t} - \frac{\partial u^b}{\partial s}
 \frac{\partial u^a}{\partial t} \}$ is the area which is surrounded by the 
small curve $\triangle C^{ab}$. We define the area element $\delta\sigma^{ab}$ 
as this small area and we obtain the area derivative of the Wilson-loop as
follows:
 \begin{eqnarray}
 \frac{\delta W[C_{xy}]}{\delta \sigma^{ab}(x)} &=& -\frac{i}{2}v_{ab}(x)
                                                W[C_{xy}]\,,\nonumber\\
 \delta\sigma^{ab} &=& \frac{1}{2} \delta s \delta t 
     \{\frac{\partial u^a}{\partial s}
       \frac{\partial u^b}{\partial t} - \frac{\partial u^b}{\partial s}
        \frac{\partial u^a}{\partial t} \} \,\,.           \label{eq.aread}
 \end{eqnarray}

The gauge transformation of the 
Wilson-loop takes the form;
 \begin{eqnarray}
 W[C_{xy}]\,\longrightarrow\,W^{\prime}[C_{xy}] = U(x)W[C_{xy}]U^{-1}(y)\,.
                                                   \label{eq.trW}
 \end{eqnarray}
In the case that the curve $C_{xy}$ is a closed loop $C_{xx}$, we see from 
(\ref{eq.trW}) 
that the operator $trW[C_{xx}]$ is gauge invariant. Since $trW[C_{xx}]$ doesn't
 depend on $x$, we may express this as $trW[C]$.  
 Note that, if we don't take the trace, $W[C_{xx}]$ 
depends on $x$. Note also that $W[C_{xx}]$ is a matrix and gauge variant 
operator although $C_{xx}$ is a closed loop.

\section{B. ~~Migdal-Makeenko equation}

Let us start with
 \begin{eqnarray}
  0 &=& \langle tr\,T^{(r)}\frac{\delta}{\delta v_a^{(r)}(x)}W[C] \rangle 
                                                             \nonumber\\
    &=& \int [dv^a]\,tr (\,T^{(r)}\frac{\delta}{\delta v_a^{(r)}(x)}
        \exp \{i\frac{1}{g^2}{\rm S}_{YM} \}W[C]\,)\,,   \label{eq.start}
 \end{eqnarray}
where the summation over the index $r$ is taken tacitly.
Here ${\rm S}_{YM}$ denotes the Yang-Mills action which is written as 
follows:
 \begin{eqnarray}
 {\rm S}_{YM} &=& \int d^4x {\cal L}_{YM}\,, \nonumber\\
 {\cal L}_{YM} &=& tr\{ -\frac{1}{16}v^{ab}v_{ab} \}\,,
 \end{eqnarray}
where ${\cal L}_{YM}$ denotes the Yang-Mills Lagrangian density. The  
path integral volume element $[dv^a]$ is normalized to satisfy the condition;
$1=\int [dv^a]\exp \{i\frac{1}{g^2}{\rm S}_{YM} \}$. The equation 
(\ref{eq.start}) is 
trivial, because it is the total derivative of the functional variable $v^a$. 
Letting the functional derivative $T^{(r)}\frac{\delta}{\delta v_a^{(r)}(x)}$ 
act on $\exp \{i{\rm S}_{YM} \}$, we find
 \begin{eqnarray}
 T^{(r)}\frac{\delta}{\delta v_a^{(r)}(x)}\exp \{i\frac{1}{g^2}{\rm S}_{YM} \}
 =  i\frac{1}{4g^2}D_b v^{ba}(x)\exp \{i\frac{1}{g^2}{\rm S}_{YM} \}\,\,,
                                                            \label{eq.dlag}
 \end{eqnarray}
where we have defined $D_b v^{ba}$ by
 \begin{eqnarray}
 D_b v^{ba}(x)= \partial_b v^{ba}(x)+\frac{i}{2}\,[v_b(x),v^{ba}(x)] \,.
 \end{eqnarray}
 eq.~(\ref{eq.start}) gives
 \begin{eqnarray}
 i\frac{1}{4g^2} \langle tr\{\,D_b v^{ba}(x)W[C_{xx}] \} \rangle
  &=& \frac{i}{2}\,\oint dw^a \langle tr\{\,T^{(r)} W[C_{xw}] T^{(r)}
 \delta^{(4)}(w-x)W[C_{wx}]\, \}\rangle\,.\nonumber\\
  & &                                       \label{eq.nontrivial}
 \end{eqnarray}
Here the $\oint$ denotes the contour integral along the closed loop $C$.

Let us try to rewrite the equation (\ref{eq.nontrivial}). Remember that the 
field strength 
$v^{ba}$ is given by the area derivative of the Wilson-loop defined in 
(\ref{eq.aread}), and the gauge connection $v_b$ by the derivative of the 
Wilson-loop defined in (\ref{eq.derivativeW}). From the equation 
(\ref{eq.aread}), we obtain
 \begin{eqnarray} 
 \frac{\delta}{\delta \sigma_{ba}(x)}\langle trW[C] \rangle = 
 -\frac{i}{2}\langle tr\{v^{ba}(x)W[C_{xx}] \} \rangle\,. \label{eq.33}
 \end{eqnarray}
Taking the derivative of the each side of the equation (\ref{eq.33}), yields
 \begin{eqnarray}
 \frac{\partial}{\partial x^b}\frac{\delta}{\delta \sigma_{ba}(x)}
 \langle trW[C] \rangle &=& -\frac{i}{2}\langle tr\{\biggl(\frac{\partial}
                            {\partial x^b}v^{ba}(x)\biggl)W[C_{xx}] \} 
                             \rangle \nonumber\\
                        & & -\frac{i}{2}\langle tr\{v^{ba}(x)
                            \biggl(\frac{\partial}{\partial x^b}W[C_{xx}]\}
                           \biggl) \rangle \,.  \label{eq.34}
 \end{eqnarray}
Recall also that $W[C_{xx}]$ depends on $x$. From the equation 
(\ref{eq.derivativeW}) we see that
 \begin{eqnarray} 
 \frac{\partial}{\partial x^b}W[C_{xx}] = -\frac{i}{2}[v_b(x),W[C_{xx}]] \,.
                                                           \label{eq.35}
 \end{eqnarray} 
Inserting (\ref{eq.35}) into (\ref{eq.34}), we find
 \begin{eqnarray}
 \frac{\partial}{\partial x^b}\frac{\delta}{\delta \sigma_{ba}(x)}
 \langle trW[C] \rangle &=&
                 -\frac{i}{2}\langle tr\{\,D_b v^{ba}(x)W[C_{xx}] \} \rangle\,.
 \end{eqnarray}
This tells us that we may express the equation (\ref{eq.nontrivial}) as 
follows:
 \begin{eqnarray}
 & &-\frac{1}{2g^2}\frac{\partial}{\partial x^b}
 \frac{\delta}{\delta \sigma_{ba}(x)}\langle trW[C] \rangle \nonumber\\
 & & =\, \frac{i}{2}\oint dw^a 
 \langle tr\{\,T^{(r)} W[C_{xw}] T^{(r)}\delta^{(4)}(w-x)W[C_{wx}]\,\}
                                        \rangle\,. \label{eq.37}
 \end{eqnarray}
 Using the property for the fundamental representation of $U(N_c)$
\beq
\sum_r T^{(r)}_{ij} \,T^{(r)}_{kl}=\delta_{il} \delta_{jk}\,, \label{eq.38}
\eeq 
 the equation (\ref{eq.37}) yields 
 \begin{eqnarray}
 & &-\frac{1}{2g^2}\frac{\partial}{\partial x^b}
 \frac{\delta}{\delta \sigma_{ba}(x)}\langle trW[C] \rangle \nonumber\\ 
 & &   =\, \frac{i}{2}\oint dw^a\delta^{(4)}(w-x) \langle tr\{\, W[C_{xw}]\,\}
         tr\{\,W[C_{wx}]\,\}\rangle  \,,  \label{eq.S-D}
 \end{eqnarray}
and this is the exact Schwinger-Dyson equation of the Wilson-loop for the 
gauge group $U(N_c)$. 
(\ref{eq.S-D}) is 
invariant under the gauge transformation. The left hand side of (\ref{eq.S-D})
 is trivially invariant because of the invariance of $trW[C]$. As for
 the right hand side, because of the delta function, the integrand is
 non-vanishing only 
if $w=x$ and, when $w=x$, each of $trW[C_{xw}]$ and $trW[C_{xw}]$ is gauge 
invariant.Therefore we see that (\ref{eq.S-D}) is gauge invariant.

Here we define the new coupling constant $g_c^2 \equiv N_c g^2$ and use 
$g_c$ instead of $g$ in the equation (\ref{eq.S-D}). 
Taking the large $N_c$ limit with $g_c$ kept 
finite, in the leading order of $N_c$ we obtain the final form;
 \begin{eqnarray}
 \frac{1}{g_c^2}\frac{\partial}{\partial x^b}
 \frac{\delta}{\delta \sigma_{ba}(x)}{\bf W}[C]
  &=& -i\,\oint dw^a\delta^{(4)}(w-x) {\bf W}[C_{xw}]{\bf W}[C_{wx}],
                                                     \label{eq.M-M}
 \end{eqnarray}
where ${\bf W}[C_{xy}]$ denotes the quantum average of the Wilson-loop;
 \[{\bf W}[C_{xy}]=\langle \frac{1}{N_c}tr\,W[C_{xy}] \rangle
 = \int [dv^a] \exp \{i\frac{N_c}{g_c^2}{\rm S}_{YM} \}
                                        \frac{1}{N_c}tr\,W[C_{xy}]\, .\]
The non-linear equation (\ref{eq.M-M}) for the ${\bf W}[C]$ is well known as 
Migdal-Makeenko equation. Note that in the leading order of 
$N_c$, ${\bf W}[C_{xy}]$ is composed of only the planar diagrams. 

In the abelian case we can easily obtain the Schwinger-Dyson equation 
which corresponds to the equation (\ref{eq.M-M});

\begin{eqnarray}
\frac{1}{g^2}\frac{\partial}{\partial x^b}
\frac{\delta}{\delta \sigma_{ba}(x)}{\bf W}[C]
 &=& -i\,{\bf W}[C]\oint dw^a\delta^{(4)}(w-x)\,\,.     \label{eq.ABM-M}
\end{eqnarray}
In this case we can calculate ${\bf W}[C]$ directly from its definition;

\[{\bf W}[C]=\langle W[C] \rangle
 = \int [dv^a] \exp \{i\frac{1}{g^2}{\rm S}_{AB} \}W[C]\, ,\]
where ${\rm S}_{AB}$ is the abelian action. We can carry out this path 
integral because of its Gaussian type. The result is 
\begin{eqnarray}
{\bf W}[C] 
&=&\exp\{-\im{2}g^2\oint\oint\,dx^a dy_a\Box^{-1}(x-y) \}\non\\
&=&\exp\{-\im{2}g^2\im{4\pi^2}\oint\oint\frac{dx^a dy_a}{(x-y)^2} \}\,.
\end{eqnarray}
This satisfies the equation (\ref{eq.ABM-M}).

\section{C. ~~Vector Superfields}

The elements of superspace are denoted by  
 \begin{eqnarray}
 z^M=(x^a,\theta^{\alpha},\bar{\theta}^{\dot{\alpha}})\,,
 \end{eqnarray}
where the capital letter $M$ represents the four-vector index $a$ as well as 
the spinor indices $\alpha$ and $\dot{\alpha}$. Elements of superspace obey 
the following multiplication law: 
\begin{eqnarray}
z^M z^N = (-)^{|M||N|}z^N z^M \,.\nonumber
\end{eqnarray}
Here $|M|$ is the following function of $M$:
\begin{eqnarray}
|M|=
{ 0, \,{\rm when\,} M\,{\rm is\,\,a\,\,vector\,\,index,} \atopwithdelims\{.
  1, \,{\rm when\,} M\,{\rm is\,\,a\,\,spinor\,\,index.}} \nonumber
\end{eqnarray}

Exterior products in superspace are defined in complete analogy to ordinary 
space:
\begin{eqnarray}
dz^M\wedge dz^N &=& -(-)^{|M||N|}dz^N\wedge dz^M \,,\nonumber\\
dz^M z^N &=&(-)^{|M||N|}z^N dz^M \,.\nonumber
\end{eqnarray}
The p-forms on superspace is expressed as
 \begin{eqnarray}
 \Omega_{\rm p}=dz^{M_1}\wedge dz^{M_2}\wedge \cdots\wedge dz^{M_{\rm p}}
       \Omega_{M_{\rm p}\cdots M_2 M_1}(z),                      \label{eq.104}
 \end{eqnarray}
where we implicitly take the summation of $M_1,M_2,\cdots,M_{\rm p}$ as in 
the usual manner for upper and lower indices. From now on we shall drop the 
symbol $\wedge$ for exterior multiplication.

We now introduce exterior derivatives on superspace which map p-forms into 
(p+1)-forms, 
 \begin{eqnarray}
 \Omega_{\rm p}&=&dz^{M_1}\cdots dz^{M_{\rm p}}
               \Omega_{M_{\rm p}\cdots M_1}(z) \,,\nonumber\\
 d\Omega_{\rm p} &=&dz^{M_1}\cdots dz^{M_{\rm p}}
        dz^M\frac{\partial}{\partial z^M}\Omega_{M_{\rm p}\cdots M_1}(z) \,.
                                                                 \label{eq.105}
 \end{eqnarray}
From this definition of exterior derivatives, we can straightforwardly show 
the following properties:
 \begin{eqnarray}
 d(\Omega_{\rm p}+\Sigma_{\rm p})
 &=&d\Omega_{\rm p}+d\Sigma_{\rm p}\,,\nonumber\\
 d(\Omega_{\rm p}\Omega_{\rm q})
 &=&\Omega_{\rm p}d\Omega_{\rm q}+(-)^{\rm q} (d\Omega_{\rm p})\Omega_{\rm q}
                                                              \,,\nonumber\\ 
 dd&=&0\,,                                                       \label{eq.106}
 \end{eqnarray}
where $\Omega_{\rm p},\,\Sigma_{\rm p}$ are p-forms and $\Omega_{\rm q}$ 
q-forms.

Let us consider a one-forms ${\cal A}$ and identify it with the gauge field on 
superspace,
 \begin{eqnarray}
 {\cal A}=dz^M{\cal A}_M(z)=dz^M{\cal A}^{(r)}_M(z)iT^r,         \label{eq.107}
 \end{eqnarray}
where $T^r$ denotes the generator of a gauge group.

 The super transformation of superfields is  
defined by using differential operators ${\rm Q}_{\alpha}$, 
$\bar{{\rm Q}}_{\dot{\alpha}}$.  
The exterior derivative $d=dz^M\frac{\partial}{\partial z^M}$ does not commute
 with ${\rm Q}_{\alpha}$, $\bar{{\rm Q}}_{\dot{\alpha}}$. Namely, 
$d=dz^M\frac{\partial}{\partial z^M}$, which maps p-forms into (p+1)-forms, 
does not maps superfields into superfields. This tells us that the basis 
$dz^M$ is not useful for supersymmetry. We will introduce a new basis in 
the following, in terms of which exterior derivatives map superfields into 
superfields. 

The following differential operators ${\rm D}_{\alpha}$, 
$\bar{{\rm D}}_{\dot{\alpha}}$ and $\partial_a$ commute or 
anticommute with ${\rm Q}_{\alpha}$, $\bar{{\rm Q}}_{\dot{\alpha}}$,
 \begin{eqnarray}
 {\rm D}_a &\equiv& \partial_a\,,\nonumber\\
 {\rm D}_{\alpha} &\equiv& \frac{\partial}{\partial\theta^{\alpha}}
  +i(\sigma^a\bta)_{\alpha}\partial_a \,,\nonumber\\
 {\rm D}_{\dot{\alpha}} &\equiv& \bar{{\rm D}}_{\dal}
  =-\frac{\partial}{\partial\bar{\theta}^{\dot{\alpha}}}
  -i(\theta\sigma^a)_{\dot{\alpha}}\partial_a \,.                \label{eq.113}
 \end{eqnarray}
We will 
express these differential operators as ${\rm D}_A$, where the capital letter
$A$ denotes the spacetime index $a$, the spinor indices $\alpha$ or 
$\dot{\alpha}$.
 The derivative $\frac{\partial}{\partial z^M}$ can be written in terms of 
${\rm D}_A$ as follows:
 \begin{eqnarray}
 \frac{\partial}{\partial z^M}=e_M\,^A {\rm D}_A\,,              \label{eq.114}
 \end{eqnarray}
where the matrix $e_M\,^A$ has the form:
 \begin{eqnarray}
 e_M\,^A=\bordermatrix{
                     &  & A & \cr
                     & \delta_a\,^b & 0 & 0 \cr
     M &-i\,_{\alpha}(\sigma^b\bar{\theta}) & \delta_{\alpha}\,^{\beta} & 0 \cr
       &-i\,_{\dot{\alpha}}(\theta\sigma^b) & 0 & -\delta_{\dot{\alpha}}
                                                         \,^{\dot{\beta}} \cr
       } \,.                                                 \label{matrix.115}
 \end{eqnarray}
We define a matrix $e_A\,^M$ such that 
\begin{eqnarray}
& &e_A\,^M e_M\,^B=\delta_A\,^B\,,\nonumber\\
& &e_M\,^A e_A\,^N=\delta_M\,^N\,.                               \label{eq.116}
\end{eqnarray}
 From this, $e_A\,^M$ has the form:
 \begin{eqnarray}
 e_A\,^M=\bordermatrix{
                     &  & M & \cr
                     & \delta_a\,^b & 0 & 0 \cr
     A & i\,_{\alpha}(\sigma^b\bar{\theta}) & \delta_{\alpha}\,^{\beta} & 0 \cr
       &-i\,_{\dot{\alpha}}(\theta\sigma^b) & 0 & -\delta_{\dot{\alpha}}
                                                         \,^{\dot{\beta}} \cr
       }\,.                                                  \label{matrix.117}
 \end{eqnarray}
By using ${\rm D}_A$ and $e_M\,^A$, the exterior derivative 
$d=dz^M\frac{\partial}{\partial z^M}$ may be written as follows:
 \begin{eqnarray}
 d=dz^M\frac{\partial}{\partial z^M}&=&dz^M e_M\,^A {\rm D}_A \nonumber\\
                                    &=&e^A {\rm D}_A \,,         \label{eq.118}
 \end{eqnarray}
where we have introduced a new basis $e^A$ defined by
 \begin{eqnarray}
 e^A=dz^M e_M\,^A\,,                                             \label{eq.119}
 \end{eqnarray}
or, from(\ref{eq.116}),
 \begin{eqnarray}
 dz^M=e^A e_A\,^M\,.                                             \label{eq.120}
 \end{eqnarray}
From now on we will call the index $M$ Einstein index, while $A$ flat index. 

We may expressed p-forms with this basis $e^A$:
 \begin{eqnarray}
 \Omega_{\rm p}&=&dz^{M_1}\cdots dz^{M_{\rm p}}
                \Omega_{M_{\rm p}\cdots M_1}(z) \,,\nonumber\\
               &=&e^{A_1}\cdots e^{A_{\rm p}}\Omega_{A_{\rm p}\cdots A_1}(z)\,.
                                                                 \label{eq.121}
 \end{eqnarray}
The relations between the components $\Omega_{M_{\rm p}\cdots M_1}$ and 
$\Omega_{A_{\rm p}\cdots A_1}$ are determined from the above equation and the 
definition of $e^A$. Since the matrix $e_M\,^A$ depends on $\theta$, 
$\bar{\theta}$, the basis $e^A$ does not commute with exterior derivatives, 
 \begin{eqnarray}
 [d,e^A]&=&dz^M e^B {\rm D}_B e_M\,^A \nonumber\\
        &=&e^C e_C\,^M e^B {\rm D}_B e_M\,^A \nonumber\\
        &=&{2i \sigma^a_{\alpha\dot{\alpha}} e^{\alpha} e^{\dot{\alpha}}
  \,\,\,(A=a) \atopwithdelims\{. 0\,\,\,\,\,\,\,({\rm otherwise})}\nonumber\\
        &\equiv& de^A\,.                                         \label{eq.122}
 \end{eqnarray}
So in terms of $e^A$, $d\Omega_{\rm p}$ takes the form, 
\begin{eqnarray}
d\Omega 
&=& e^{A_1}\cdots e^{A_{\rm p}} e^B {\rm D}_B \Omega_{A_{\rm p}\cdots A_1}
                                                        \nonumber\\
& & + e^{A_1}\cdots (de^{A_{\rm p}})\Omega_{A_{\rm p}\cdots A_1}\nonumber\\
& & + \cdots\cdots  \nonumber\\
& & + (de^{A_1})\cdots e^{A_{\rm p}}\Omega_{A_{\rm p}\cdots A_1}\,.
\end{eqnarray}
Note that $de^A$ is expressed only with $e^{\alpha}$, $e^{\dot{\alpha}}$ and 
does not depends on $z=(x,\theta,\bar{\theta})$. Namely, in terms of $e^A$, 
each component of $d\Omega_{\rm p}$ is represented with ${\rm D}_A$ and the 
components $\Omega_{A_{\rm p}\cdots A_1}$ of $\Omega_{\rm p}$. This tells us 
that, in terms of $e^A$, if all components of $\Omega_{\rm p}$ are 
superfields, all components of $d\Omega_{\rm p}$ are also superfields.
(Note that, in general, when $\Omega_{A_{\rm p}\cdots A_1}$ are superfields, 
$\Omega_{M_{\rm p}\cdots M_1}$ are not superfields.)
From the above argument, we can conclude that the basis $e^A$ is useful for 
supersymmetry.

Let us represent the one-forms ${\cal A}$ by using the basis $e^A$, 
 \begin{eqnarray}
 {\cal A}&=&dz^M {\cal A}_M \nonumber\\
         &=&e^A {\cal A}_A\,,
 \end{eqnarray}
and from this, the relations between ${\cal A}_M$ and ${\cal A}_A$ are 
\begin{eqnarray}
{\cal A}_{\underline{a}}
&=&{\cal A}_a \,,\nonumber\\
{\cal A}_{\underline{\alpha}}
&=&{\cal A}_{\alpha}-i(\sigma^a\bar{\theta})_{\alpha}{\cal A}_a
                                                             \,,\nonumber\\
{\cal A}_{\underline{\dot\alpha}}
&=&-{\cal A}_{\dot\alpha}-i(\theta\sigma^a)_{\dot\alpha}
{\cal A}_a\,, \label{eq.125}
\end{eqnarray}
where $\underline{a}$, $\underline{\alpha}$ and $\underline{\dal}$ 
denote Einstein indices. Here we regard all the components ${\cal A}_A$ as 
superfields. The gauge transformation of the component ${\cal A}_A$ takes the 
form, 
\begin{eqnarray}
{\cal A}_A\longrightarrow{\cal A}^{\prime}_A
                =-X^{-1}{\rm D}_A X+X^{-1}{\cal A}_A X.    \label{eq.126}
\end{eqnarray}

The field strength ${\cal F}$ may be rewritten in terms of $e^A$ as follows:
 \begin{eqnarray}
 {\cal F}
 &=&d{\cal A}+{\cal A}{\cal A} \nonumber\\
 &=&e^Ae^B{\rm D}_B{\cal A}_A+(de^A){\cal A}_A \nonumber\\
 & &+e^A{\cal A}_Ae^B{\cal A}_B \nonumber\\
 &=&2i\, e\sigma^a\bar{e} {\cal A}_a \nonumber\\
 & &+\frac{1}{2}e^Ae^B\{{\rm D}_B{\cal A}_A-(-)^{|A||B|}{\rm D}_A{\cal A}_B
                                                          \nonumber\\
 & &\hspace*{1.7cm} -{\cal A}_B{\cal A}_A+(-)^{|A||B|}{\cal A}_A{\cal A}_B\}\,,
 \end{eqnarray}
and components of ${\cal F}$ are 
\begin{eqnarray}
{\cal F}_{ab}
&=&\partial_a{\cal A}_b-\partial_b{\cal A}_a-[{\cal A}_a,{\cal A}_b]
                                                            \,,\nonumber\\
{\cal F}_{a \alpha}
&=&\partial_a{\cal A}_{\alpha}-{\rm D}_{\alpha}{\cal A}_a
   -[{\cal A}_a,{\cal A}_{\alpha}] \,,\nonumber\\
{\cal F}_{a \dot{\alpha}}
&=&\partial_a\bar{{\cal A}}_{\dot{\alpha}}
   -\bar{{\rm D}}_{\dot{\alpha}}{\cal A}_a
   -[{\cal A}_a,\bar{{\cal A}}_{\dot{\alpha}}] \,,\nonumber\\
{\cal F}_{\alpha\beta}
&=&{\rm D}_{\alpha}{\cal A}_{\beta}+{\rm D}_{\beta}{\cal A}_{\alpha}
   -\{{\cal A}_{\alpha},{\cal A}_{\beta} \} \,,\nonumber\\
{\cal F}_{\dot{\alpha}\dot{\beta}}
&=&\bar{{\rm D}}_{\dot{\alpha}}\bar{{\cal A}}_{\dot{\beta}}
   +\bar{{\rm D}}_{\dot{\beta}}\bar{{\cal A}}_{\dot{\alpha}}
   -\{\bar{{\cal A}}_{\dot{\alpha}},\bar{{\cal A}}_{\dot{\beta}} \}
                                                  \,,\nonumber\\
{\cal F}_{\alpha\dot{\alpha}}
&=&{\rm D}_{\alpha}\bar{{\cal A}}_{\dot{\alpha}}
   +\bar{{\rm D}}_{\dot{\alpha}}{\cal A}_{\alpha}
   -\{{\cal A}_{\alpha},\bar{{\cal A}}_{\dot{\alpha}} \} 
   +2i\sigma^a_{\alpha\dot{\alpha}} {\cal A}_a \,,               \label{eq.128}
\end{eqnarray}
where we define these by ${\cal F}=\frac{1}{2}e^Ae^B{\cal F}_{BA}$.
Each component of ${\cal F}$ represents a full superfield multiplet. These 
multiplets contain a large number of component fields. Most of the component 
fields are superfluous and must be eliminated through constraint equations. 
The constraint equations which we should take are well-known as flatness 
condition:
\begin{eqnarray}
{\cal F}_{\alpha\beta}={\cal F}_{\dot{\alpha}\dot{\beta}}
={\cal F}_{\alpha\dot{\alpha}}=0 \,.                             \label{eq.129}
\end{eqnarray}
Since our goal in this appendix is not to get the general solutions of 
(\ref{eq.129}), here we adopt a following solution \cite{So,GJ}
\begin{eqnarray}
{\cal A}_{\alpha}
&=&-{\rm e}^{-{\rm V}}{\rm D}_{\alpha}{\rm e}^{\rm V} \,,\nonumber\\
\bar{{\cal A}}_{\dot{\alpha}}&=&0 \,,\nonumber\\
{\cal A}_a
&=&-\frac{i}{4}\bar{\sigma}^{\dot{\alpha}\alpha}_a
\Bigl({\rm D}_{\alpha}\bar{{\cal A}}_{\dot{\alpha}}
   +\bar{{\rm D}}_{\dot{\alpha}}{\cal A}_{\alpha}
   -\{{\cal A}_{\alpha},\bar{{\cal A}}_{\dot{\alpha}} \} \Bigr) \nonumber\\
&=&\frac{i}{4}\bar{\sigma}^{\dot{\alpha}\alpha}_a
   \bar{{\rm D}}_{\dot{\alpha}}{\rm e}^{-{\rm V}}{\rm D}_{\alpha}
                                            {\rm e}^{\rm V} \,,\label{eq.sol}
\end{eqnarray}
where ${\rm V}={\rm V}^{(r)}T^r$ and ${\rm e}^{\rm V}$ is an element of the 
gauge group generated by $T^r$.  
These equations trivially satisfy 
${\cal F}_{\alpha\beta}={\cal F}_{\dot{\alpha}\dot{\beta}}=0$, because 
 $\bar{{\cal A}}_{\dot{\alpha}}=0$ and 
${\cal A}_{\alpha}$ has the form of pure gauge type.
 ${\cal A}_a$ is determined to satisfy 
${\cal F}_{\alpha\dot{\alpha}}=0$ for the given 
$\bar{{\cal A}}_{\dot{\alpha}}$, ${\cal A}_{\alpha}$.

We will assume that ${\rm V}$ is a real superfield;${\rm V}={\rm V}^{\dagger}$.
${\rm V}$ has the following form:
\begin{eqnarray}
{\rm V}
&=&{\rm C}(x)+i\theta\chi(x)+i\bar{\theta}\bar{\chi}(x) \nonumber\\ 
& &+\frac{i}{2}\theta\theta[{\rm M}(x)+i{\rm N}(x)]
   -\frac{i}{2}\bar{\theta}\bar{\theta}[{\rm M}(x)-i{\rm N}(x)] \nonumber\\
& &-\theta\sigma^a\bar{\theta}v_a(x)
   +i\theta\theta\bar{\theta}[\bar{\lambda}(x)
                      +\frac{i}{2}\bar{\sigma}^a\partial_a\chi(x)] \nonumber\\
& &-i\bar{\theta}\bar{\theta}\theta[\lambda(x)
                      +\frac{i}{2}\sigma^a\partial_a\bar{\chi}(x)] \nonumber\\
& &+\frac{1}{2}\theta\theta\bar{\theta}\bar{\theta}
             [{\rm D}(x)+\frac{1}{2}\Box{\rm C}(x)] \,,       \label{eq.vector}
\end{eqnarray}
where the component fields ${\rm C},{\rm D},{\rm M},{\rm N}$ and $v_a$ are all 
real from the reality condition of ${\rm V}$.
${\rm V}$ is called vector superfield and is a fundamental superfield in 
supersymmetric gauge theories. Since we have kept explicitly supergauge 
covariance in these arguments, the solutions which are given by exchanging 
$e^{\rm V}$ for $e^{\rm V}X$ in (\ref{eq.sol}) are also satisfy the flatness 
conditions. ${\rm V}$ has another ambiguity. Even if we exchange $e^{\rm V}$ 
for $e^{\Lambda^{\dagger}}e^{\rm V}$ where $\Lambda^{\dagger}$ satisfies 
${\rm D}_{\alpha}\Lambda^{\dagger}=0$, the equations (\ref{eq.sol}) does not 
change. Combining this ambiguity with the gauge transformation in superspace, 
the transformation of ${\rm V}$ takes the form 
$e^{\rm V}\to e^{\rm V^{\prime}}=e^{\Lambda^{\dagger}}e^{\rm V}X$. Since 
${\rm V^{\prime}}$ should also be real, $X$ must be equal to $e^{\Lambda}$. 
Therefore ${\rm V}$ transforms as follows:
\begin{eqnarray}
e^{\rm V}&\longrightarrow& 
e^{\rm V^{\prime}}=e^{\Lambda^{\dagger}}e^{\rm V}e^{\Lambda} \,,\nonumber\\ 
& &\bar{\rm D}_{\dot{\alpha}}\Lambda=0 \,,\nonumber\\
& &{\rm D}_{\alpha}\Lambda^{\dagger}=0 \,.                  \label{eq.sgaugetr}
\end{eqnarray}
We will call this transformation the restricted super gauge transformation in 
this paper because the element $X$ of super gauge must satisfy 
$X={\rm e}^{\Lambda},\, \bar{\rm D}_{\dal}\Lambda=0$ in the above 
transformation.(note that, in the case $X={\rm e}^{\Lambda}$, 
${\cal A}_{\dal}$ remains zero, 
${\cal A}_{\dal}\to {\cal A}^{\prime}_{\dal}=-{\rm e}^{-\Lambda}{\rm D}_{\dal}
{\rm e}^{\Lambda}=0$.)
%
%

Let us define a new superfield ${\cal W}_{\alpha}$,
\begin{eqnarray}
{\cal W}_{\alpha}
&=&\frac{i}{2}\sigma_{\alpha\dot{\alpha}}^a\bar{\rm D}^{\dot{\alpha}}{\cal A}_a
                                                           \nonumber\\ 
&=&-\frac{1}{4}\bar{\rm D}\bar{\rm D}
       {\rm e}^{-{\rm V}}{\rm D}_{\alpha}{\rm e}^{\rm V} \,. \label{eq.defW}
\end{eqnarray}
In terms of $(y,\theta,\bar{\theta})$, $\bar{\rm D}_{\dot{\alpha}}
=-\frac{\partial}{\partial\bar{\theta}^{\dot{\alpha}}}$. From this we see that 
\begin{eqnarray}
\bar{\rm D}_{\dot{\alpha}}\bar{\rm D}\bar{\rm D}
=-\frac{\partial}{\partial\bar{\theta}^{\dot{\alpha}}}
 \frac{\partial}{\partial\bar{\theta}^{\dot{\beta}}}
 \frac{\partial}{\partial\bar{\theta}_{\dot{\beta}}}=0  \,,\nonumber
\end{eqnarray}
and, furthermore,
\begin{eqnarray}
\bar{\rm D}_{\dot{\alpha}}{\cal W}_{\alpha}=0\,.
\end{eqnarray}
Namely ${\cal W}_{\alpha}$ is a chiral superfield. The transformation of 
${\cal W}_{\alpha}$ which follows (\ref{eq.sgaugetr}) takes the form,
\begin{eqnarray}
{\cal W}_{\alpha}\longrightarrow{\cal W}_{\alpha}^{\prime}
=e^{-\Lambda}{\cal W}_{\alpha}e^{\Lambda}\,.                \label{eq.135}
\end{eqnarray}\


\begin{thebibliography}{99}

\bibitem{MM}
  Yu.M. Makeenko and A.A. Migdal,  {\sl Phys. Lett.}~{\bf B88}~(1979)~135.

\bibitem{Mig}
  A.A. Migdal, {\sl Phys. Rep.}~{\bf 102}~(1983)~199.

\bibitem{Pol}
  A.M. Polyakov, {\sl ``Contemporary Concepts in Physics Volume 3; 
  Gauge Fields and Strings''},~~{\sl harwood academic publishers}.

\bibitem{IT}
   H.Itoyama and H.Takashino, {\sl Phys. lett.}~{\bf B381}~(1996)163.

 \bibitem{Gates}
 S. J. Gates, {\sl Phys. Rev.}~{\bf D16}~(1977)~1727.

\bibitem{Mar}
  S.~Marculescu, {\sl Nucl. Phys.}~{\bf B213}~(1983)~523.

\bibitem{GGRS}
   S. J. Gates, M. Grisaru, M. Rocek and W. Siegel,~
 {\sl ''Superspace''}~{\sl Benjamin/Cummings}~(1983).

\bibitem{WB}
  J. Wess and J. Bagger,
 {\sl  ``Supersymmetry and Supergravity''},~~{\sl Princeton University
 Press}~(1992) 2nd edition.

\bibitem{So} 
  M.~F.~Sohnius, {\sl Phys. Rep.}~{\bf 128}~(1985)~39.

\bibitem{GJ}
   J-L. Gervais, M.T. Jaekel and A. Neveu,
 {\sl Nucl. Phys.}~{\bf B155}~(1979)~75.

\bibitem{MakMed}
  Yu.M. Makeenko and P.B. Medvedev, {\sl Nucl. Phys.}~{\bf B193}~(1981)~444.

\bibitem{IH}
  J.Ishida, A.Hosoya, {\sl Prog. Theor. Phys.}~{\bf 62}~(1979)544.

\bibitem{GerNe}
  J. L. Gervais, A. Neveu, {\sl Nucl. Phys.}~{\bf B163}~(1980)189 .

\bibitem{Are}
  I. Aref'eva, {\sl Phys. Lett.}~{\bf 93B}~(1980)347.

\bibitem{DV}
  V. S. Dotsenko, S. N. Vergeles, {\sl Nucl. Phys.}~{\bf B169}~(1980)527.

\bibitem{CD}
  N. S. Cragie, H. Dorn, {\sl Nucl. Phys.}~{\bf B185}~(1981)204.

\bibitem{Ao}
  S. Aoyama, {\sl Nucl. Phys.}~{\bf B194}~(1982)513.

\bibitem{SW}
   N.~Seiberg and E.~Witten, {\sl Nucl. Phys.}~{\bf B426}~(1994)19:
 {\sl Nucl. Phys.}~{\bf B431}~(1994)484.  See K. Intriligator and N. Seiberg
 {\sl Nucl. Phys. Proc. Suppl.}~{\bf 45BC}~(1996)1,hep-th 9509066  
 for more references.

\bibitem{Int}
 A.Gorsky,I.Krichever,A.Marshakov,A.Mironov
 and  A.Morozov {\sl Phys. Lett.}~{\bf 355B}~(1995)466, hep-th/9505035;
 E.Martinec and N.Warner, {\sl Nucl. Phys.}~{\bf B459}~(1996)97,hep-th/9509161;
 T.Nakatsu and K.Takasaki, {\sl Mod. Phys. Lett.}~{\bf A11}~(1996)157,
 hep-th/9509162; R.Donagi and E.Witten, {\sl Nucl. Phys.}~{\bf B460}~(1996)299,
 hep-th/9510101; T.Eguchi and S.K.Yang, hep-th/9510183; 
 E.Martinec, {\sl Phys. Lett.}~{\bf B 367}~(1996)91,
 hep-th/9510204; H.Itoyama and A.Morozov, 
 {\sl Nucl. Phys.}~{\bf B477}~(1996)855,hep-th/9511126; 
 hep-th/9512161,{\sl Nucl. Phys.}~{\bf B488}~[{\rm PM}] to appear.

\end{thebibliography}
\end{document}